\documentclass[useAMS,usenatbib]{mn2e}

\usepackage{natbib}
\usepackage{graphicx}
\usepackage{amssymb}
\usepackage{color}
\usepackage{caption}
\usepackage{lipsum,graphicx,multicol}
\usepackage{float}

\usepackage{subcaption}

\title[Variations on a theme of AGN outflows]{Variations on a theme of AGN-driven outflows: \\ luminosity evolution and ambient density distribution}
\author[ ]
{W. Ishibashi$^{1,}$$^{2}$\thanks{E-mail: wako@ast.cam.ac.uk} and A. C. Fabian$^{1}$
\footnotemark[0]\\
\footnotemark[0]\\
$^{1}$Institute of Astronomy, Madingley Road, Cambridge CB3 0HA \\
$^{2}$Physik-Institut, Universitat Zurich, Winterthurerstrasse 190, 8057 Zurich, Switzerland 
}

\begin{document}

\pdfminorversion=4

\date{Accepted ? Received ?; in original form ? }

\pagerange{\pageref{firstpage}--\pageref{lastpage}} \pubyear{2012}

\maketitle

\label{firstpage}

\begin{abstract} 
Galactic outflows are now commonly observed in starburst and active galactic nuclei (AGN) host galaxies. Yet, there is no clear consensus on their physical driving mechanism(s). We have previously shown that AGN radiative feedback, driven by radiation pressure on dust, can account for the observed dynamics and energetics of galactic outflows, provided that radiation trapping is taken into account. Here we generalise our model results by explicitly considering the temporal evolution of the central AGN luminosity,  and the shell mass evolution in different ambient density distributions. In the case of fixed-mass shells, the high observed values of the momentum ratio ($\zeta = \dot{p}/(L/c)$) and energy ratio ($\epsilon_k = \dot{E}_{k}/L$) may be attributed to either radiation trapping or AGN luminosity decay. In contrast, for expanding shells sweeping up mass from the surrounding environment, a decay in AGN luminosity cannot account for the observed high energetics, and radiation trapping is necessarily required. 
Indeed, strong radiation trapping, e.g. due to high dust-to-gas ratios, can considerably boost the outflow energetics. 
We obtain a distinct radial dependence for the outflow energetics ($\zeta(r)$, $\epsilon_k(r)$) in the case of radiation trapping and luminosity decay, which may help discriminate between the two scenarios. In this framework, the recently discovered `fossil' outflows, with anomalously high values of the energetics, may be interpreted as relics of past AGN activity. The observed outflow properties may therefore provide useful constraints on the past history of AGN activity and/or the physical conditions of the outflow launch region. 
\end{abstract}



\section{Introduction}

Powerful outflows on galactic scales are now starting to be commonly observed. Such high-velocity ($v \sim 1000$km/s) outflows are found to extend on $\sim$kpc scales, and are detected in different gas phases \citep{Sturm_et_2011, Veilleux_et_2013, Cicone_et_2014, Gonzalez-Alfonso_et_2017, Fiore_et_2017}. 
Molecular outflows are of particular interest, since they carry the bulk of the outflowing mass and constitute the medium from which stars eventually form. Observations of molecular outflows indicate high mass outflow rates ($\dot{M} \gtrsim 10^2 M_{\odot}/yr$), high momentum flux ($\dot{p} \gtrsim 10 L/c$), and high kinetic power ($\dot{E}_k \gtrsim 0.01 L$), with the quoted values usually reported for outflow radii of $r \lesssim 1$kpc \citep{Cicone_et_2014, Fiore_et_2017}.

Despite the remarkable observational progress, the physical origin of galactic outflows and their driving mechanism(s) are still much debated. In some cases, it is not even clear whether the central source is a nuclear starburst or the active galactic nucleus (AGN). In particular, when powerful outflows are detected on $\sim$kpc scales in a galaxy without any sign of ongoing AGN activity at the centre, the observed outflows are assumed to be starburst-driven. Indeed, a number of high-velocity outflows have been attributed to compact starbursts, without the need to invoke AGN feedback \citep{Diamond-Stanic_et_2012, Sell_et_2014, Geach_et_2014}. 
Alternatively, `fossil' outflows powered by a past AGN event (which has since faded) may persist to the present day \citep[e.g.][]{King_2010}. 

Concerning the outflow driving mechanisms, two main physical models have been discussed in the literature: wind shocks and radiation pressure. Within the shocked wind scenario, the energy-driven outflows are predicted to have large momentum rates ($\dot{p} \sim 20 L/c$) and high kinetic luminosities ($\dot{E}_k \sim 0.05 L$), consistent with observations \citep{Zubovas_King_2012, Faucher-Giguere_Quataert_2012}. Radiation pressure on dust is another way of driving large-scale feedback, based on the enhanced radiation-matter coupling, due to the large dust absorption cross section \citep{Fabian_1999, Murray_et_2005, Thompson_et_2015}. 

The high energetics observed in galactic outflows initially seemed to favour wind energy-driving over radiation pressure-driving.
However, we have recently shown that AGN radiation pressure on dust is capable of accounting for the observed outflow energetics, provided that radiation trapping is properly taken into account \citep{Ishibashi_et_2018}. Moreover, the observational scaling relations with luminosity (a sub-linear scaling for the mass outflow rate, $\dot{M} \propto L^{1/2}$, and a super-linear scaling for the kinetic power, $\dot{E}_k \propto L^{3/2}$) are naturally obtained in our framework. 
We note that in our previous work, we assumed the simple case of a constant central luminosity and fixed-mass shell approximation, which allowed us to derive analytic limits and gain some physical insight into the problem.

However, in more realistic situations, the AGN luminosity is likely to vary in time, and the outflowing shell may sweep up mass from the surrounding environment. Here we generalise our model results by explicitly considering the temporal evolution of the central AGN luminosity and the shell mass evolution in different ambient density distributions. 

The paper is structured as follows. In Section 2, we briefly recall the basics of AGN radiation pressure-driven outflows and the associated energetics. In Section 3, we consider different histories for the temporal evolution of the AGN luminosity; while in Section 4, we consider the sweeping-up of mass from different ambient density distributions. The combined effects of AGN luminosity and shell mass evolutions are analysed in Section 5, and the case of extreme radiation trapping is considered in Section 6. 
We discuss the resulting physical implications in Section 7.


\section{AGN radiation pressure-driven outflows}

As in our previous work, we consider AGN feedback driven by radiation pressure on dust. 
We recall that the equation of motion of the radiation pressure-driven shell is given by:
\begin{equation}
\frac{d}{dt} [M_{sh}(r) v] = \frac{L(t)}{c} (1 + \tau_{IR} - e^{-\tau_{UV}} ) - \frac{G M(r) M_{sh}(r)}{r^2}
\end{equation}
where $L(t)$ is the central luminosity, $M(r)$ is the total mass distribution, and $M_{sh}(r)$ is the shell mass \citep{Thompson_et_2015, Ishibashi_Fabian_2015, Ishibashi_et_2018}. 
We assume an isothermal potential ($M(r) = \frac{2 \sigma^2}{G} r$, where $\sigma$ is the velocity dispersion), and explicitly consider the temporal evolution of the AGN luminosity, $L(t)$, as well as the radial evolution of the shell mass, $M_{sh}(r)$. 
The infrared (IR) and ultraviolet (UV) optical depths are given by:
\begin{equation}
\tau_{IR}(r) = \frac{\kappa_{IR} M_{sh}(r)}{4 \pi r^2} 
\end{equation}
\begin{equation}
\tau_{UV}(r) = \frac{\kappa_{UV} M_{sh}(r)}{4 \pi r^2} 
\end{equation}
where $\kappa_{IR}$=$5 \, \mathrm{cm^2 g^{-1} f_{dg, MW}}$ and $\kappa_{UV}$=$10^3 \, \mathrm{cm^2 g^{-1} f_{dg, MW}}$ are the IR and UV opacities, with the dust-to-gas ratio normalised to the Milky Way value.
We assume the following values as fiducial parameters of the model, unless specified otherwise: $L = 3 \times 10^{46}$erg/s, $M_\mathrm{sh,fix} = 10^8 M_{\odot}$, $R_0 = 50$pc, and $\sigma$ = 200 km/s.

The observed outflows are characterised by the mass outflow rate ($\dot{M}$), the momentum flux ($\dot{p}$), and the kinetic power ($\dot{E}_k$). By analogy with the observational works \citep[e.g.][]{Gonzalez-Alfonso_et_2017}, the three physical quantities are defined as:
\begin{equation}
\dot{M}(r) = \frac{M_{sh}(r)}{t_{flow}} = M_{sh}(r) \frac{v}{r}
\label{Eq_Mdot}
\end{equation} 
\begin{equation}
\dot{p}(r)  = \dot{M} v= M_{sh}(r) \frac{v^2}{r}
\label{Eq_pdot}
\end{equation} 
\begin{equation}
\dot{E}_k(r) = \frac{1}{2} \dot{M} v^2 = \frac{1}{2} M_{sh}(r) \frac{v^3}{r} 
\label{Eq_Ekdot}
\end{equation}

Two derived quantities are often used in quantifying the outflow energetics, the momentum ratio ($\zeta$) and the energy ratio ($\epsilon_k$), respectively given by: 
\begin{equation}
\zeta(r) = \frac{\dot{p}}{L/c} = M_{sh}(r) \frac{v^2}{r} \frac{c}{L}
\label{momentum_ratio}
\end{equation} 
\begin{equation}
\epsilon_k(r) = \frac{\dot{E}_k}{L}  = \frac{1}{2} M_{sh}(r) \frac{v^3}{r} \frac{1}{L}
\label{energy_ratio}
\end{equation}

Observations of molecular outflows typically indicate energetics in the range: $\dot{M} \sim (10^2-10^3) M_{\odot}/yr$, $\dot{p} \sim (1-30) L/c$, $\dot{E}_k \sim (0.3-3) \%$ \citep[from][and references therein]{Fiore_et_2017}. 
We recall that there is a difference by a factor of $\sim 3$, when assuming spherical uniform density vs. thin shell-like geometry \citep{Maiolino_et_2012, Gonzalez-Alfonso_et_2017}.


\section{Luminosity temporal evolution}
\label{luminosity_evolution}

\begin{figure*}
\begin{multicols}{3}
    \includegraphics[width=\linewidth]{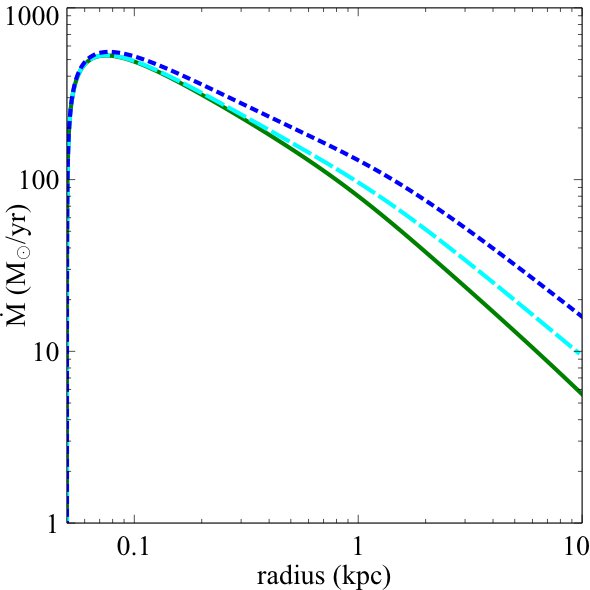}\par
    \includegraphics[width=\linewidth]{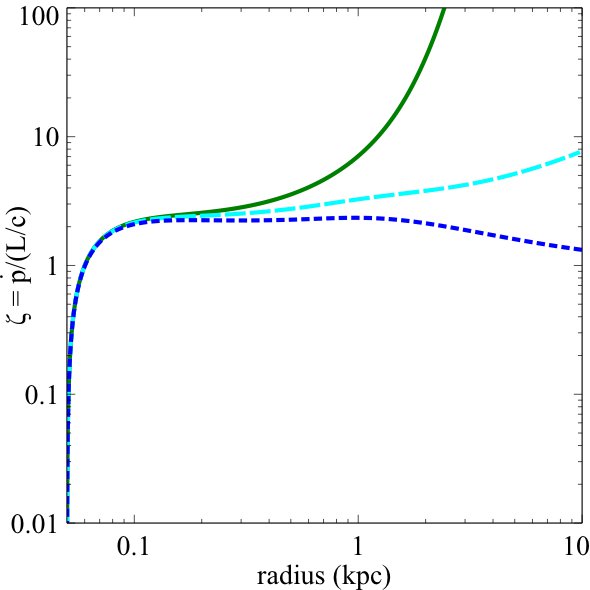}\par 
    \includegraphics[width=\linewidth]{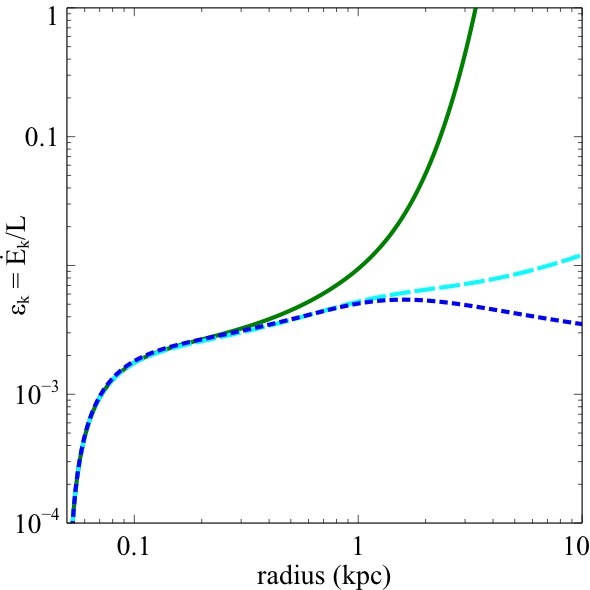}\par 
    \end{multicols}
\caption{Energetics of fixed-mass shells ($M_{sh}(r) = 10^8 M_{\odot}$, \textbf{$\tau_{\mathrm{IR,0}} = 3.5$}) for different AGN luminosity decays ($L_0 = 3 \times 10^{46}$erg/s): exponential decay with $t_c = 5 \times 10^5$yr (green solid), power-law decay with $t_d = 10^6$yr and $\delta = 2$ (cyan dashed), power-law decay with $t_d = 10^6$yr and $\delta = 1$ (blue dotted).
Left panel: mass outflow rate versus radius; middle panel: momentum ratio versus radius; right panel: energy ratio versus radius.}
\label{plot_energetics_fixedmass}
\end{figure*}

The observational measurements of the outflow parameters are used to compute the momentum ratio and the energy ratio, as defined in Eqs. (\ref{momentum_ratio}-\ref{energy_ratio}). Galactic outflows are usually detected at a given distance from the centre (e.g. $r \sim 1$kpc), while the central luminosity is derived from the current observed flux. However, a certain amount of time is required for the outflow to reach the present location; and over this time span, the central luminosity is unlikely to have stayed constant. Indeed, AGNs are known to be variable objects, and their luminosities can vary by several orders of magnitude over a range of timescales. This can have a significant impact on the inferred values of the outflow energetics, as we already suggested \citep[][see also \citet{Zubovas_2018}]{Ishibashi_Fabian_2015}. 

We first consider the temporal evolution of the central luminosity, $L(t)$, in the case of fixed-mass shells. A decay in AGN luminosity can be expected, as the amount of accreting gas decreases at late times, possibly following the disruption of the fuel supply by the radiation pressure-driven outflow. 
If the accretion rate abruptly falls off, we may expect a rapid decay in luminosity (e.g. exponential decay); while a more gentle decay (e.g. power-law decay) may be expected if the accretion disc slowly dissipates on a viscous timescale. 

We thus consider different forms for the central AGN luminosity decay, the exponential decay:
\begin{equation}
L(t) = L_0 e^{-t/t_c} \, , 
\end{equation}
where $L_0$ is the initial luminosity and $t_c$ is a characteristic decay timescale; and the power-law decay: 
\begin{equation}
L(t) = L_0 (1 + t/t_d)^{-\delta} \, , 
\end{equation} 
where $\delta$ is the power-law slope and $t_d$ is a characteristic timescale.

In Figure \ref{plot_energetics_fixedmass}, we plot the mass outflow rate, the momentum ratio, and the energy ratio, as a function of radius for different AGN luminosity decays. We recall that the outflow continues to propagate, despite the drop in central luminosity, with faster luminosity decays leading to lower shell velocities. Thus at a given radius, the mass outflow rate is lowest for the most rapid luminosity decay, i.e. the exponential decay. 
A decrease in AGN luminosity implies higher momentum and energy ratios, and values comparable to the observed ones are attained on $\sim$kpc scales.

From Figure \ref{plot_energetics_fixedmass}, we see that the exponential luminosity decay leads to very high momentum and energy ratios ($\zeta \sim 100$ and $\epsilon_k \sim 1$) at large radii. Such extreme values are usually not observed in galactic outflows, and thus an exponential luminosity decay may not be favoured in general (but see the case of fossil outflows). 
On the other hand, power-law luminosity decays lead to much less variations in the inferred energetics, such that the momentum and energy ratio stay in the range $\zeta \sim (1-10)$ and $\epsilon_k \sim (10^{-3}-0.01)$, consistent with the typically observed values \citep{Fiore_et_2017}. 
In this picture, short characteristic decay timescales ($< 10^6$yr) are required in order to obtain high values of the momentum and energy ratios at small radii ($r \lesssim 1$kpc), compatible with the observational measurements.

We have previously shown that radiation trapping can account for the high observed values of the momentum and energy ratios, which are naturally achieved on small scales, where the outflow is optically thick to the reprocessed radiation \citep{Ishibashi_et_2018}. In the case of fixed-mass shells, here we see that high momentum and energy ratios may also be attributed to a rapid luminosity decay, provided that the characteristic timescales are quite short (without the need to invoke strong radiation trapping). 
We note that the AGN temporal variability argument is most relevant in interpreting sources currently accreting at significantly sub-Eddington rates, which form a major fraction of the observational samples \citep{Veilleux_et_2013, Cicone_et_2014, Fluetsch_et_2018}.


\section{Shell mass evolution}
\label{Shell_mass_evolution}

\begin{figure*}
\begin{multicols}{3}
    \includegraphics[width=\linewidth]{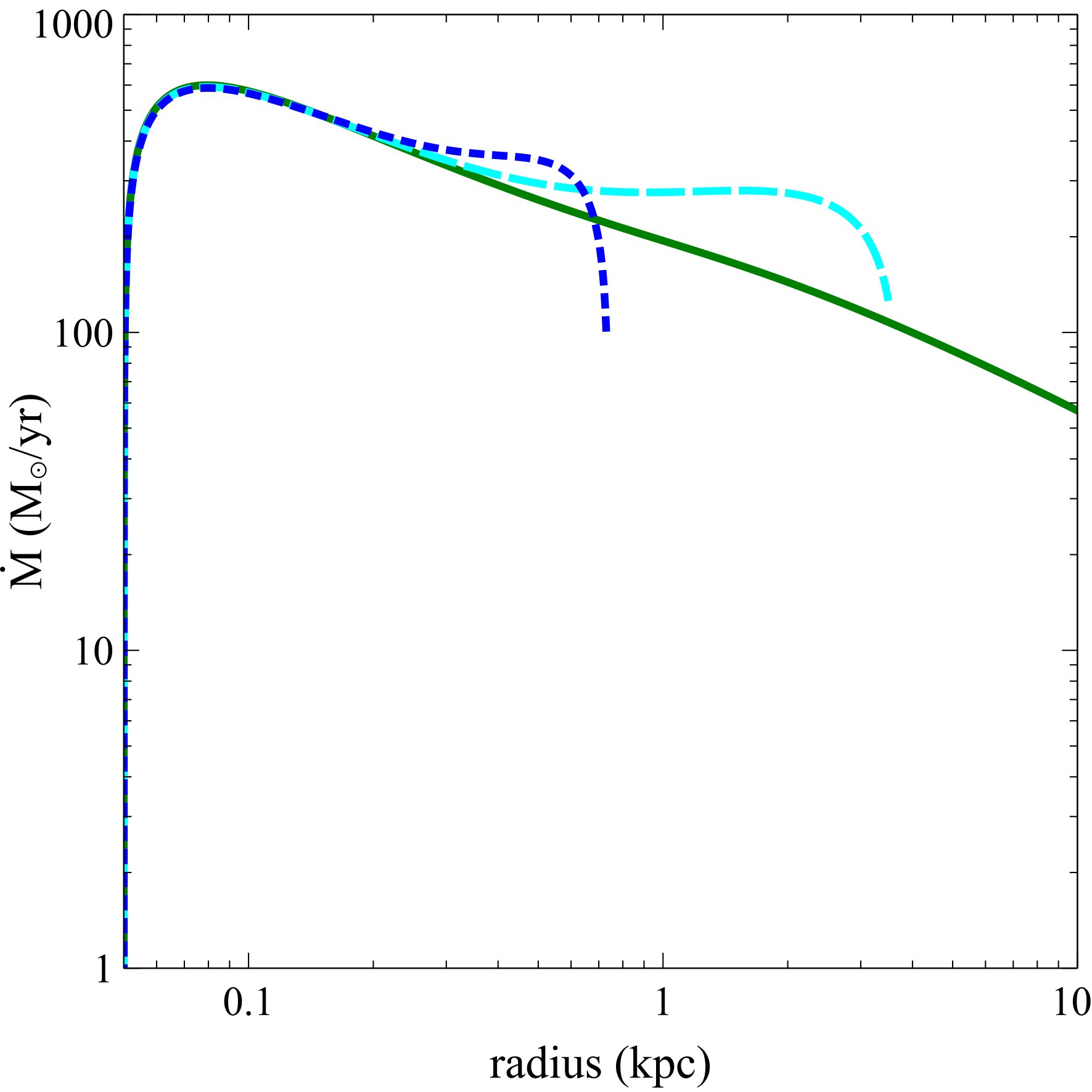}\par
    \includegraphics[width=\linewidth]{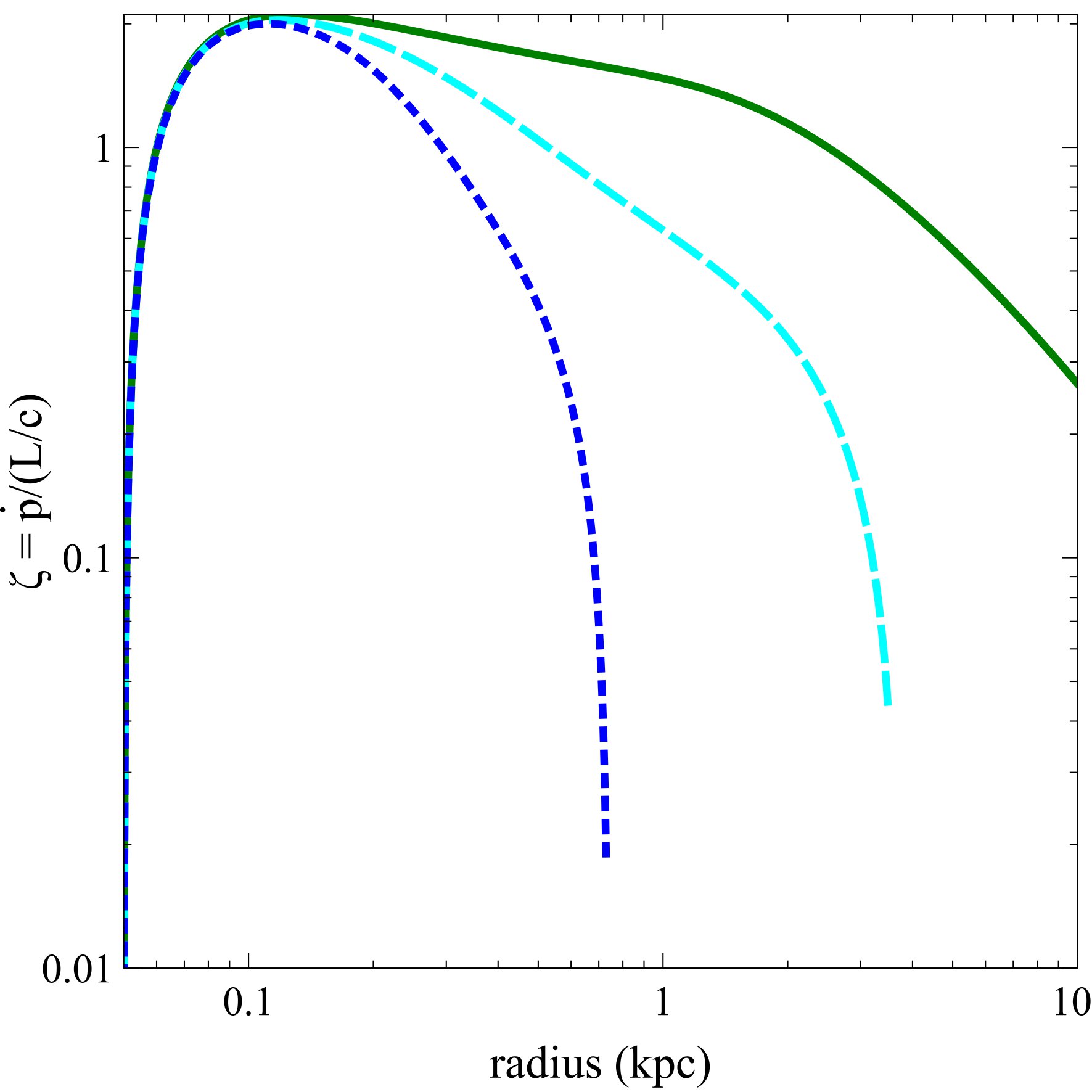}\par 
    \includegraphics[width=\linewidth]{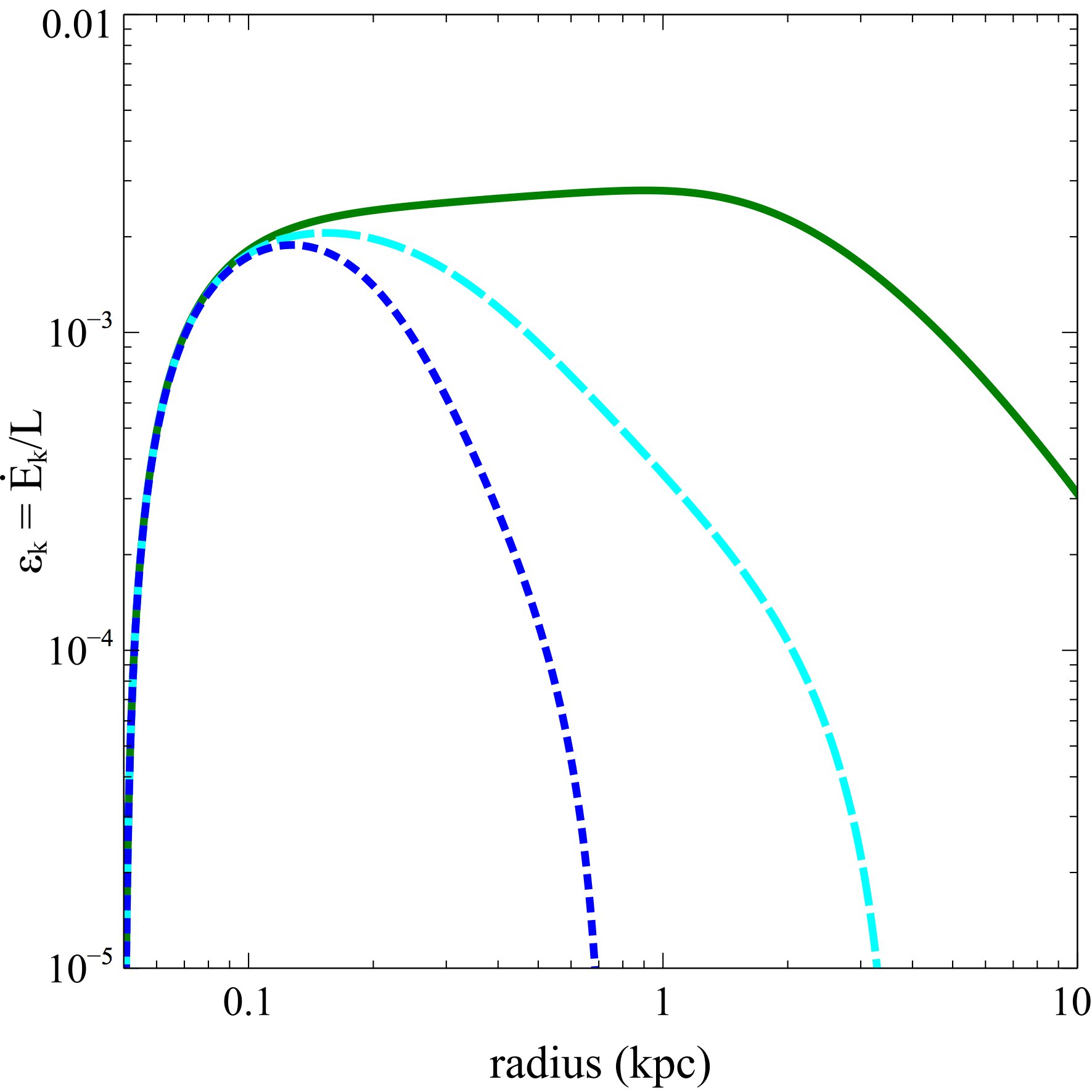}\par 
    \end{multicols}
\caption{Energetics of expanding shells with constant luminosity ($L(t) = L_0 =  3 \times 10^{46}$erg/s), $M_0 = 10^8 M_{\odot}$, and $n_0 = 100 cm^{-3}$: constant density distribution with $\alpha = 0$ and $\tau_{\mathrm{IR,0}} = 3.6$ (blue dotted); power-law density distribution with $\alpha = 1$ and $\tau_{\mathrm{IR,0}} = 3.6$ (cyan dashed); isothermal density distribution with $\alpha = 2$ and $\tau_{\mathrm{IR,0}} = 3.7$ (green solid). 
Left panel: mass outflow rate versus radius; middle panel: momentum ratio versus radius; right panel: energy ratio versus radius.}
\label{plot_energetics_expanding}
\end{figure*}

While thin shell structures are well justified in some cases \citep[][and references therein]{Thompson_et_2015}, more in general, the outflowing shell is likely to sweep up mass from the surrounding environment, as it expands into the host galaxy. 
We consider the case of expanding shells sweeping up mass from the ambient density distribution, parametrised as a power-law of radius: 
\begin{equation}
n(r) = n_0 \left( \frac{r}{R_0} \right)^{-\alpha}
\end{equation}
where $\alpha$ is the power-law exponent, $n_0$ is the density of the external medium, and $R_0$ is the initial radius. 
The shell mass is then given by:
\begin{equation}
M_{sh}(r) = M_0 + 4 \pi m_p \int n(r) r^2 dr \approx M_0 + 4 \pi m_p n_0 R_0^{\alpha} \frac{r^{3-\alpha}}{3-\alpha}
\end{equation} 
where $M_0$ is the initial mass of the shell at $R_0$. 
Here we consider the effects of different power-law exponents ($\alpha = 0, 1, 2$), while keeping the central luminosity constant ($L(t) = L_0$). We note that $\alpha = 0$ corresponds to a constant density distribution, and $\alpha = 2$ corresponds to an isothermal density distribution. We focus here on variations in the slope of the ambient density distribution, while the effects of the external density $n_0$ and the initial radius $R_0$ are analysed in Appendix \ref{App_A1}. 

Figure \ref{plot_energetics_expanding} shows the mass outflow rate, the momentum ratio, and the energy ratio, as a function of radius for different ambient density distributions. 
As the expanding shells sweep up mass from the surrounding environment, they are continuously decelerated, with the shell velocities decreasing more rapidly for flatter density distributions. For a constant density distribution ($\alpha = 0$), the swept-up mass increases steeply with radius ($M_{sh}(r) \propto r^3$), while for an isothermal distribution ($\alpha = 2$), it rises more slowly following a linear relation ($M_{sh}(r) \propto r$). 
As a result of the sweeping-up of ambient material, the expanding shell velocities rapidly fall off, and high values of the outflow energetics can never be reached, irrespective of the precise value of the power-law exponent. 
In particular, the momentum ratio and the energy ratio always remain around $\zeta \sim 1$ and $\epsilon_k \sim 10^{-3}$, respectively.  
Therefore the high values of the energetics observed in galactic outflows cannot be reproduced in the case of expanding shells sweeping up ambient material, regardless of the exact ambient density distribution.

In the $\alpha = 0$ case, the IR optical depth grows linearly with radius ($\tau_{IR} \propto r$), but even this is not enough to launch large-scale outflows, as the swept-up mass increases more steeply with radius ($M_{sh}(r) \propto r^3$). As we shall see later, an increase in the dust-to-gas ratio may temporarily help the outflow propagation, but the shell will eventually fall back. 
In any case, a constant density distribution ($\alpha = 0$) is not the most physically plausible case.


\section{Combination: luminosity and mass evolutions}

\begin{figure*}
\begin{multicols}{3}
    \includegraphics[width=\linewidth]{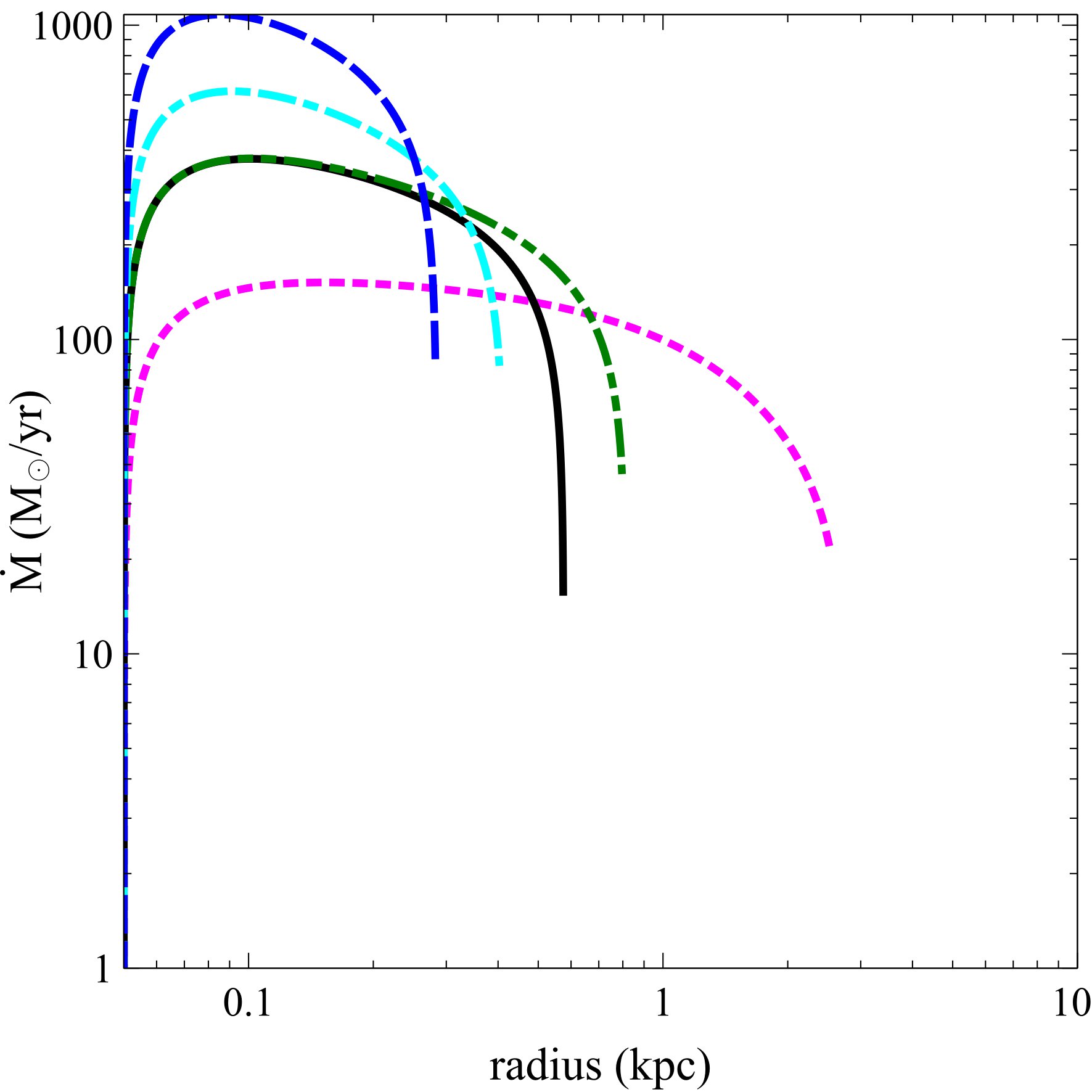}\par
    \includegraphics[width=\linewidth]{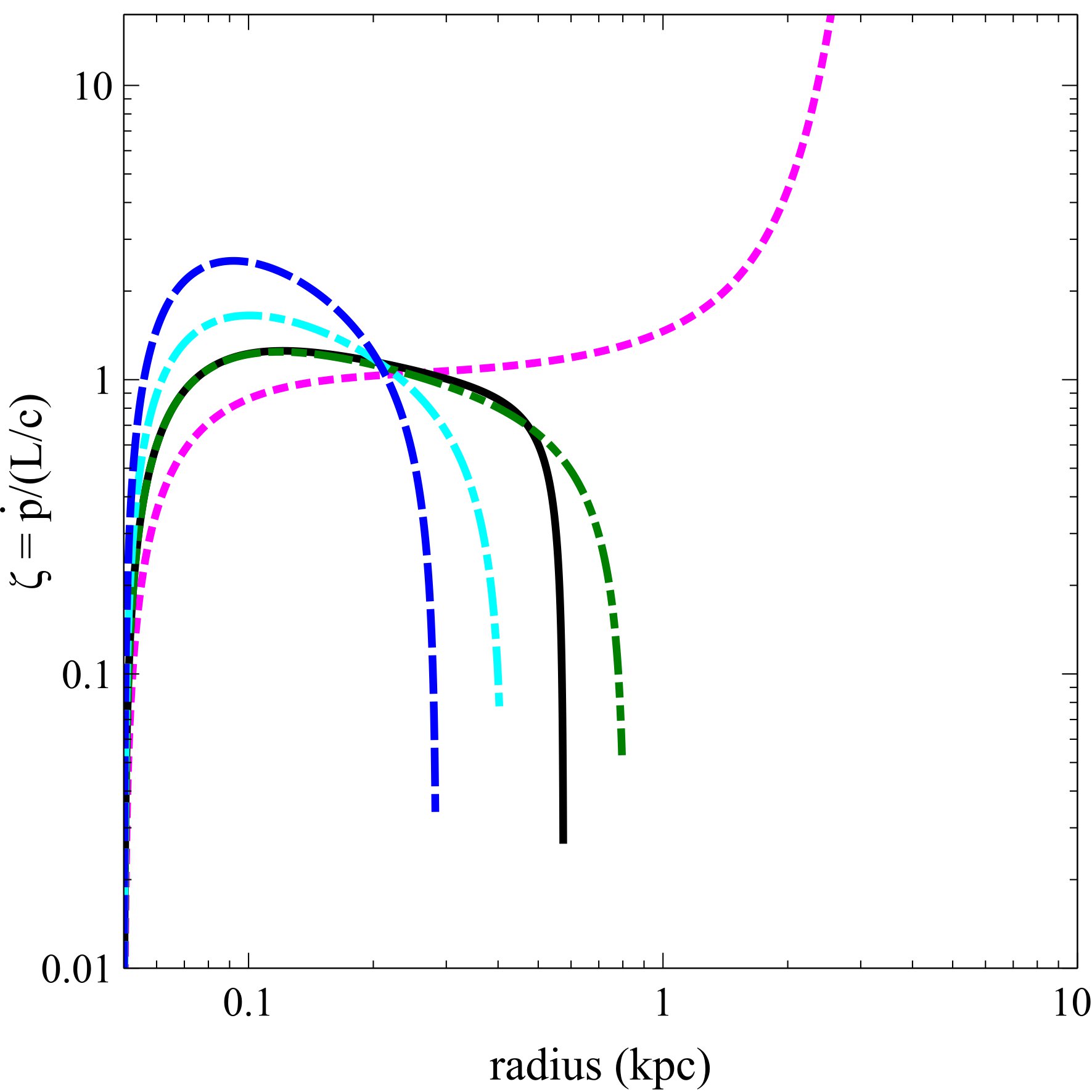}\par 
    \includegraphics[width=\linewidth]{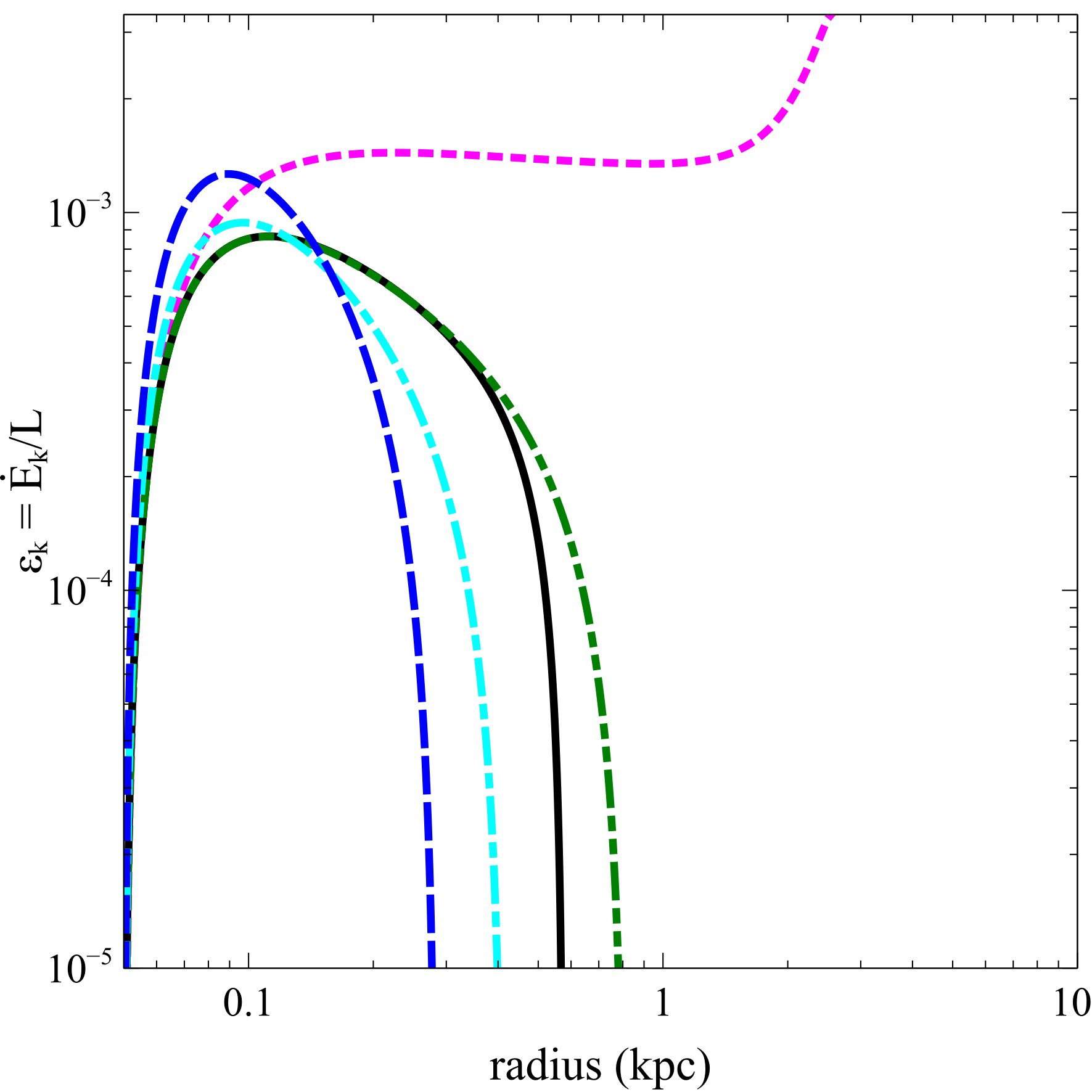}\par 
    \end{multicols}
\caption{Energetics of isothermal shells with power-law decay ($L_0 = 3 \times 10^{46}$erg/s, $\delta = 1$, $t_d = 10^6$yr) and different gas fractions: $f = 0.2$ and $\tau_{\mathrm{IR,0}} = 6.4$ (blue dashed), $f = 0.1$ and $\tau_{\mathrm{IR,0}} = 3.2$ (cyan dash-dot), $f = 0.05$ and $\tau_{\mathrm{IR,0}} = 1.6$ (green dash-dot-dot); with exponential luminosity decay ($L_0 = 3 \times 10^{46}$erg/s, $t_c = 10^6$yr) and different gas fractions: $f = 0.05$ and $\tau_{\mathrm{IR,0}} = 1.6$ (black solid), $f = 0.01$ and $\tau_{\mathrm{IR,0}} = 0.3$ (magenta dotted). }
\label{plot_energetics_isothermal}
\end{figure*}

We now combine the temporal evolution of the AGN luminosity and the shell mass evolution. 
Here we consider the case of expanding shells sweeping up mass from an isothermal density distribution (`isothermal shell'), which should be a reasonable approximation in the inner regions of the galaxy. The corresponding shell mass is given by: $M_{sh}(r) = \frac{2 f \sigma^2}{G} r$, where $f$ is the gas fraction. 
We follow the isothermal shell evolution for a power-law decay in AGN luminosity, as well as for an exponential luminosity decay (which previously led to the highest inferred momentum and energy ratios, see Sect. \ref{luminosity_evolution}). 

Figure \ref{plot_energetics_isothermal} shows the resulting mass outflow rate, momentum ratio, and energy ratio, as a function of radius for isothermal shells with different values of the gas fraction. Due to a combination of moderate gas fractions and fast luminosity decay, the outflowing shells do not reach very high speeds. We see that for typical gas fractions of $f \sim 0.1$, isothermal shells never reach high values of the momentum and energy ratios, for both power-law and exponential luminosity decays. 
In particular, we observe that the momentum ratio and energy ratio always remain $\zeta \lesssim$ a few  and $\epsilon_k \lesssim 10^{-3}$ (for gas fractions in the range $f \sim 0.05-0.2)$, even for an exponential luminosity decay. This is in strong contrast to the case of fixed-mass shells, for which very high momentum and energy ratios are usually obtained with the exponential decay (cf. Sect. \ref{luminosity_evolution}). 

In the case of isothermal shells, high values of the outflow energetics may only be obtained for very low gas fractions ($f \sim 0.01$, magenta dotted curve). However, this is only a temporary increase, since both the momentum ratio and energy ratio will eventually drop at larger radii, where the shell velocity tends towards zero. In fact, there is no `critical' value of the gas fraction below which the outflow energetics can grow indefinitely. A decrease in the gas fraction just leads to a later drop of the outflow energetics, occurring at larger radii.

We observe that for isothermal shells, a drop in AGN luminosity alone cannot account for the high values of the momentum and energy ratios observed in galactic outflows, irrespective of the precise history of the AGN luminosity decay. 
This suggests that a considerable degree of radiation trapping ($\tau_{IR} \gtrsim 10$) is instead required.
In fact, for an isothermal shell with $f \sim 0.1$ and Galactic dust-to-gas ratio, the corresponding IR optical depth is typically of the order of $\tau_{\mathrm{IR,0}} \sim 3$. 
Therefore, in the general case of expanding shells sweeping up mass from an isothermal distribution (with plausible gas fractions, $f \sim 0.1$), the observed high outflow energetics cannot be explained by AGN temporal variability, and hence must be attributed to radiation trapping.


\section{Extreme radiation trapping}

\begin{figure*}
\begin{multicols}{2}
    \includegraphics[width=0.8\linewidth]{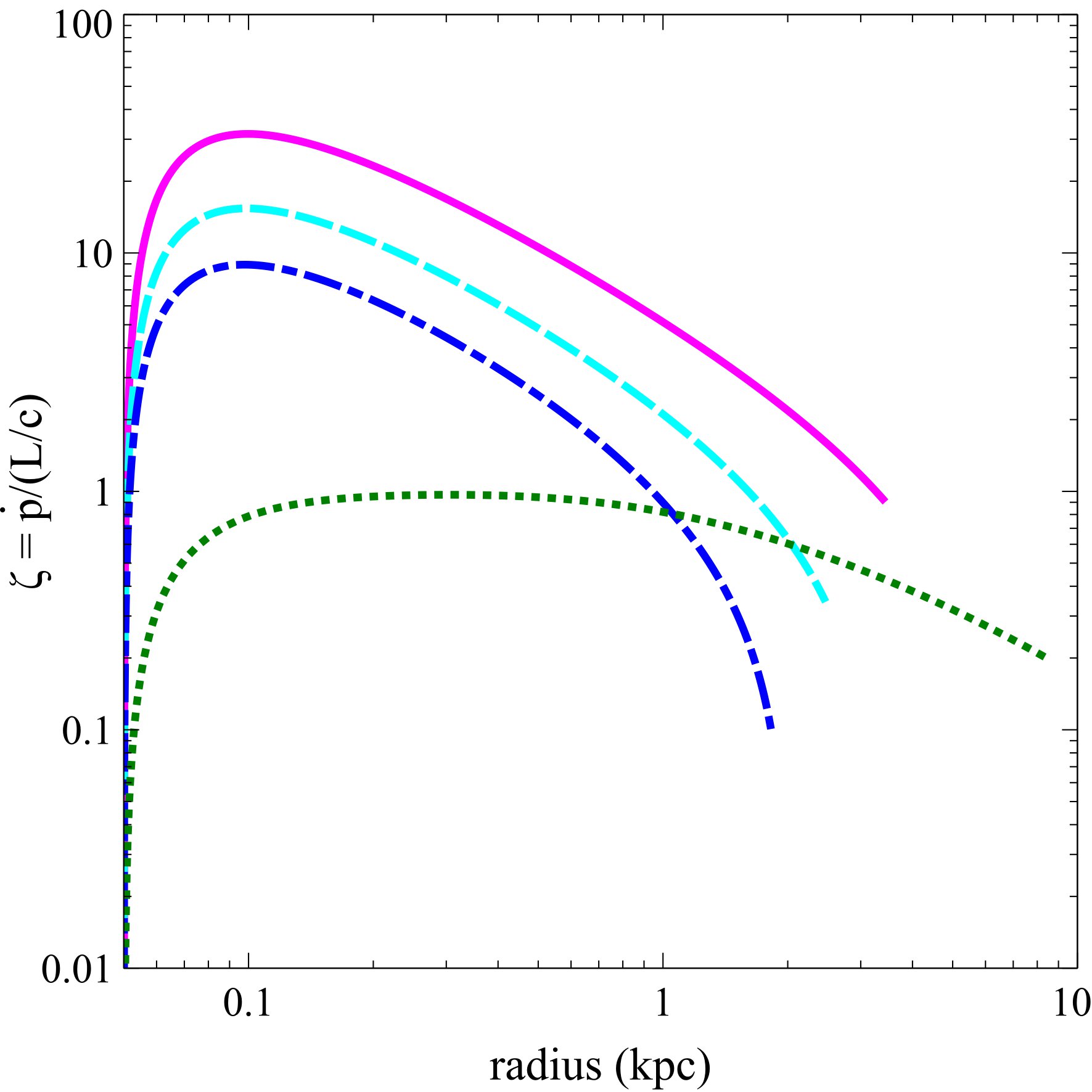}\par
    \includegraphics[width=0.8\linewidth]{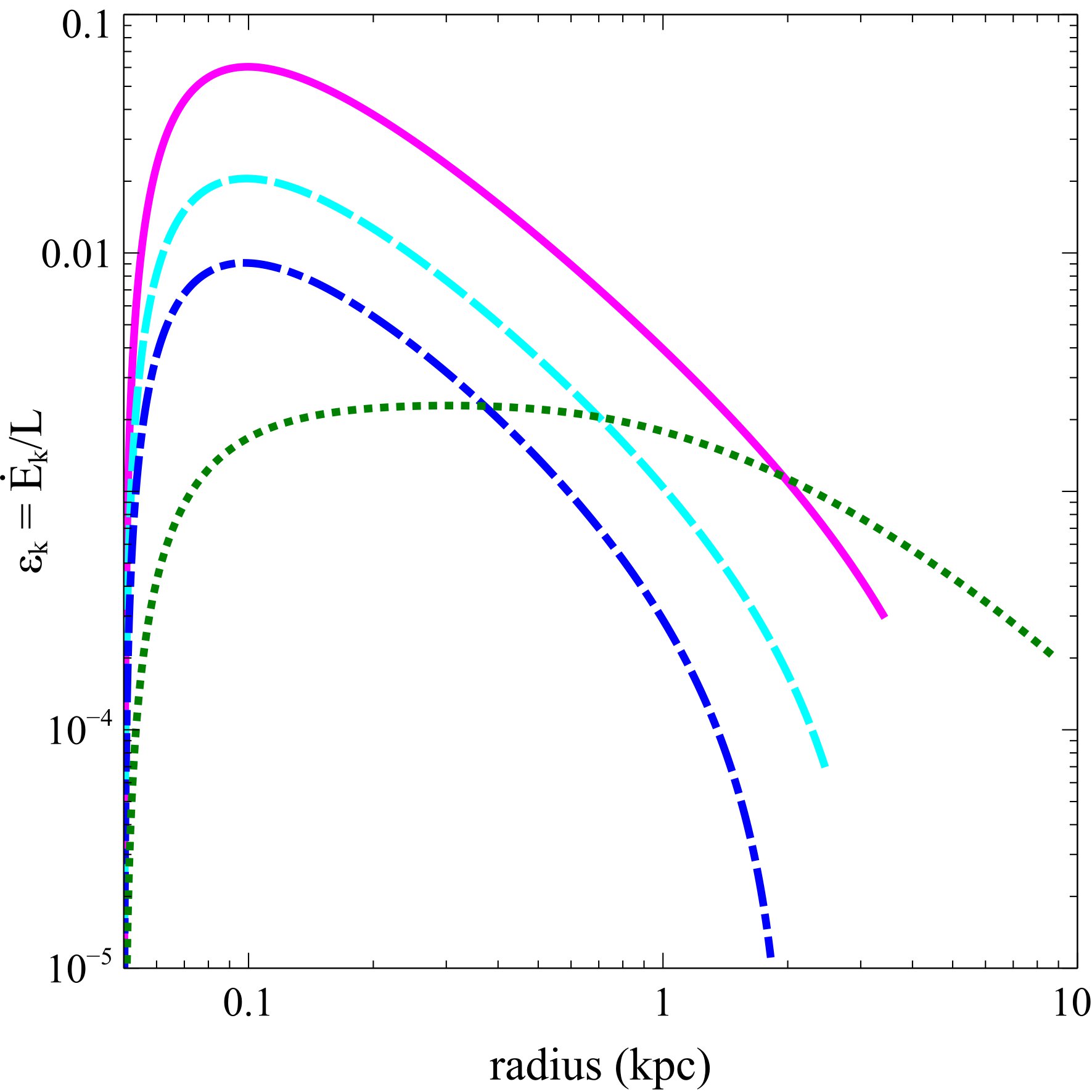}\par 
    \end{multicols}
\caption{Momentum ratio (left) and energy ratio (right) of isothermal shells ($f = 0.2$) with constant luminosity ($L = 3 \times 10^{46}$erg/s) and variations in the dust-to-gas ratio: $f_{dg} = 1/50$ and $\tau_{\mathrm{IR,0}} = 19.1$ (blue dash-dotted), $f_{dg} = 1/30$ and $\tau_{\mathrm{IR,0}} = 31.8$ (cyan dashed), $f_{dg} = 1/15$ and $\tau_{\mathrm{IR,0}} = 63.6$ (magenta solid). For comparison, the isothermal shell  of Fig. \ref{plot_energetics_expanding} (with $f = 0.004$ and $f_{dg} = 1/150$, $\tau_{\mathrm{IR,0}} = 0.13$) is also plotted (green dotted).
}
\label{plot_VarFdg_re}
\end{figure*}

A possible way of boosting the outflow energetics is via the trapping of reprocessed radiation. The required conditions of extreme optical depths are likely reached in the nuclear regions of ULIRG-like systems.
Following e.g. a merger event, large amounts of gas are funnelled into the inner $\sim 100$pc region. Due to the angular momentum barrier, the infalling gas cannot be immediately accreted by the central black hole, and accumulates in the nucleus, leading to nuclear starbursts. 
The combination of very high column densities and large dust concentration observed in the innermost regions of ULIRGs \citep[e.g.][and references therein]{Aalto_et_2015} may provide the required conditions for efficient radiation trapping. 

Observational results indicate that large amounts of dust can be produced in core-collapse supernovae \citep{Gomez_et_2012, Owen_Barlow_2015, Wesson_et_2015}. The resulting dust-to-gas ratio in the supernova remnants can be quite high, e.g. of the order of $\sim 1/30$ in the case of the Crab Nebula \citep{Owen_Barlow_2015}. These observations also favour dust grains with large sizes, which are more likely to survive destruction by shock sputtering. 
Moreover, grain growth in the interstellar medium is likely to contribute to the increase of the dust mass at later times. Large dust masses are observed in submillimeter galaxies at high redshifts, with associated high dust-to-gas ratios of $\sim 1/50$ \citep{Kovacs_et_2006, Michalowski_et_2010}. Recent ALMA observations of dust-reddened quasars indicate a range of dust-to-gas ratios, with values up to $\sim 1/30$ \citep{Banerji_et_2017}.
Since the dust-to-gas ratio is also known to correlate with metallicity, high $f_{dg}$ values are naturally expected in the central regions of dense starbursts \citep[][and references therein]{Andrews_Thompson_2011}.

Nuclear starbursts are characterised by high star formation rates, and associated high supernova rates, which should efficiently contribute to the dust enrichment of the local environment. 
As a consequence, the assumption of a typical Milky-Way dust-to-gas ratio ($f_{dg} \sim 1/150$) may not be appropriate in ULIRG-like systems, which comprise the majority of the observational samples \citep[e.g.][]{Cicone_et_2014}.
Recent observations of extreme molecular outflows in two local ULIRGs suggest heavily dust-obscured nuclei, with very high dust-to-gas ratios, $f_{dg} \sim 1/20-1/10$ (although this could be partly attributed to the degeneracy between dust optical depth and dust temperature) \citep{Gowardhan_et_2018}.

Here, we explore the effects of the dust-to-gas ratio on the outflow energetics by probing a range of $f_{dg}$ values. 
In Figure \ref{plot_VarFdg_re}, we plot the momentum ratio and energy ratio for isothermal shells with enhanced dust-to-gas ratios. As the IR opacity directly scales with the dust-to-gas ratio, an enhancement in $f_{dg}$ eventually leads to a higher IR optical depth, which facilitates the trapping of radiation. 
In contrast to the plots in the previous sections, we observe that high values of the momentum ratio and energy ratio ($\zeta \gtrsim 10$ and $\epsilon_k \gtrsim 1\%$) can now be obtained. 
We also note that the energy ratios get an even greater boost than the momentum ratios. 
Therefore, in the case of expanding shells sweeping up mass, radiation trapping is a necessary ingredient for reproducing the high energetics of observed outflows. This further reinforces the importance of radiation trapping within the AGN radiative feedback scenario.

Extreme radiation trapping will also affect the observed spectral energy distribution, as the optical/UV emission may be completely absorbed by dust, and re-emitted in the IR band. 
In the initial stages, the central source is thus heavily enshrouded and the AGN may be hidden from sight. 
An observational example of an outflow episode, just starting to emerge, is given by the Hot DOG W2246-0526, which is observed close to the point of blowing out its dusty cocoon in a large-scale outflow \citep{Diaz-Santos_et_2016}. 
At later times, when the radiation pressure-driven outflow has cleared out the obscuring dusty gas from the inner regions and reached the galactic scales, we may observe a bright optical/UV quasar at the centre. 
For instance, the molecular outflow in Mrk 231 has a size of $\gtrsim 1$kpc \citep{Feruglio_et_2015}, and may have been launched a few million years ago. We have previously discussed how AGN radiative feedback may provide a natural physical interpretation for the observed co-evolutionary path, from dust-obscured starbursts to unobscured luminous quasars \citep{Ishibashi_Fabian_2016b}, and how our models may also be applied to the recently discovered populations of dusty quasars \citep{Ishibashi_et_2017}.


\section{Discussion}
\label{discussion}

\subsection{Luminosity decay or radiation trapping?}

\begin{figure*}
\begin{multicols}{2}
    \includegraphics[width=0.8\linewidth]{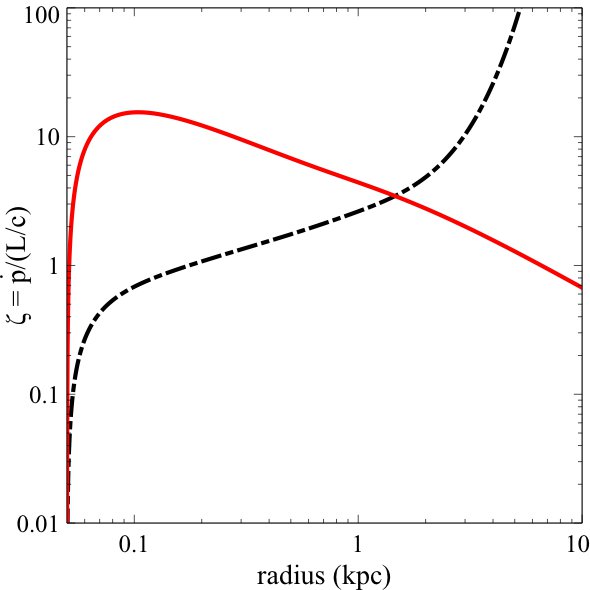}\par
    \includegraphics[width=0.8\linewidth]{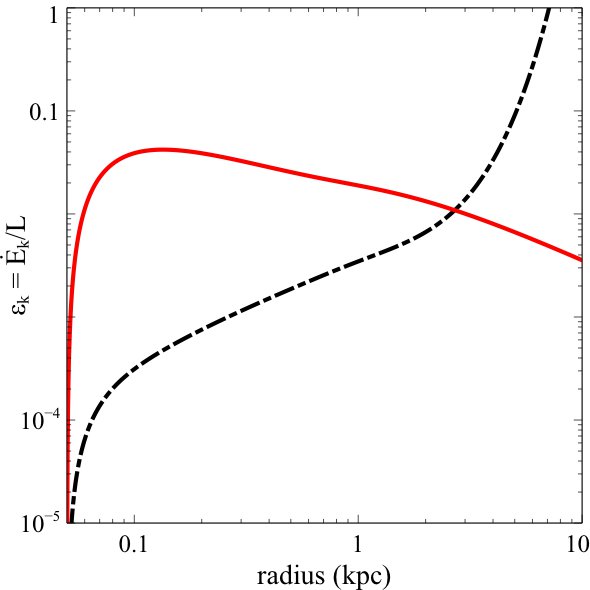}\par 
    \end{multicols}
\caption{Radial profile of the momentum ratio (left) and energy ratio (right) of fixed-mass shells for strong radiation trapping vs. exponential luminosity decay: $L = 3 \times 10^{46}$erg/s, $\tau_{IR,0} = 30$ (red solid); $L_0 = 3 \times 10^{46}$erg/s, $t_c = 10^6$yr, $\tau_{IR,0} = 1$ (black dash-dot). 
}
\label{plot_VarTrap_fix_re}
\end{figure*}
\begin{figure*}
\begin{multicols}{2}
    \includegraphics[width=0.8\linewidth]{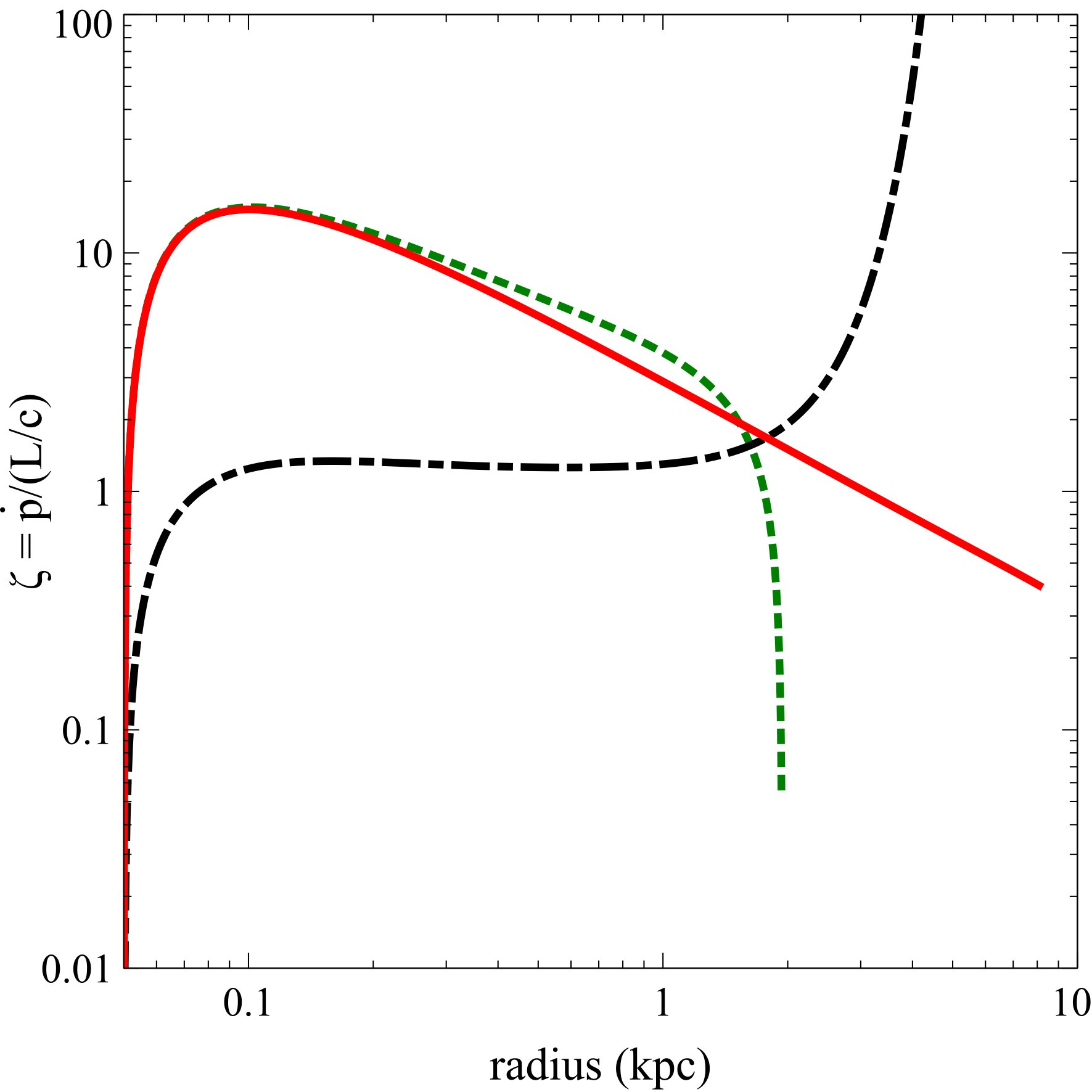}\par
    \includegraphics[width=0.8\linewidth]{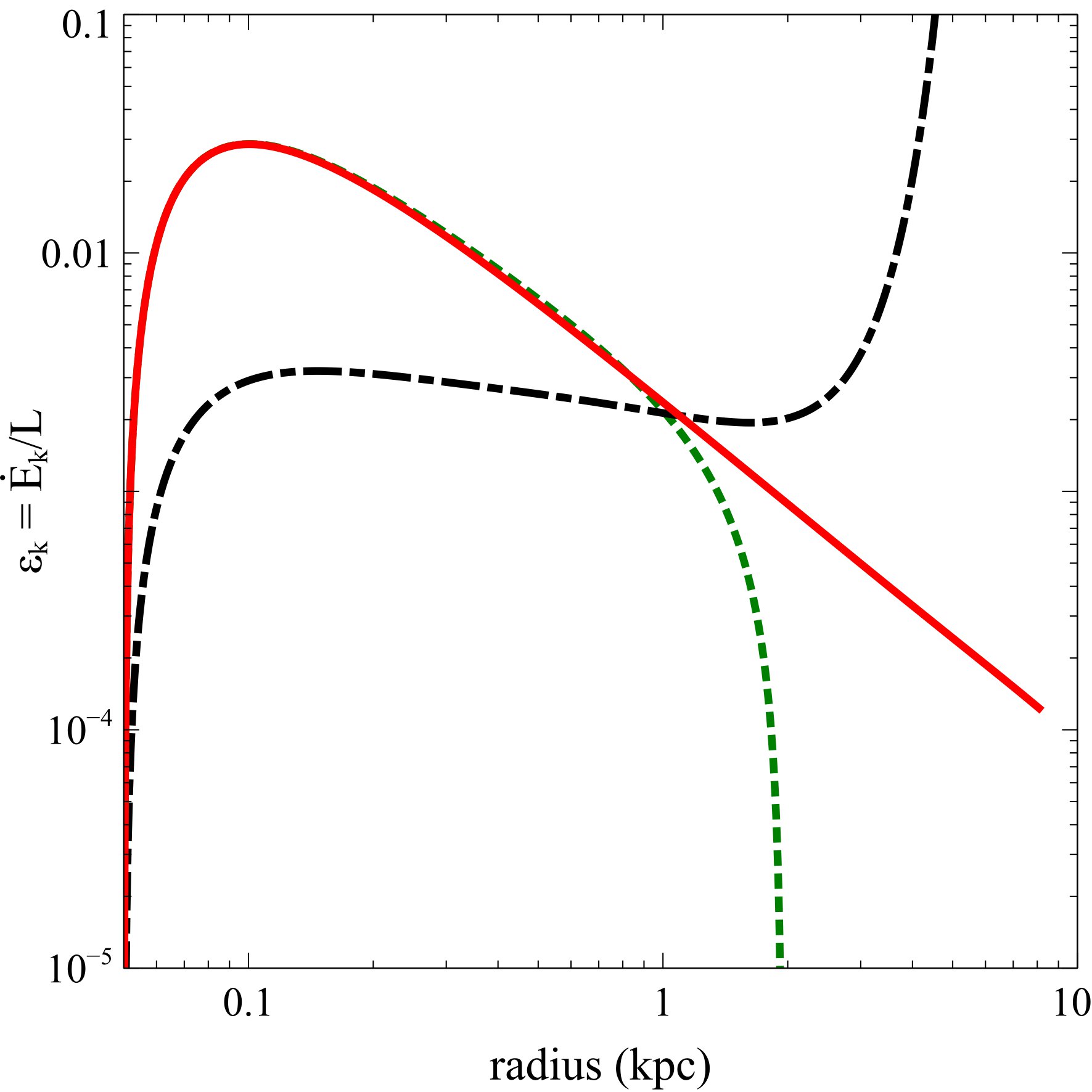}\par 
    \end{multicols}
\caption{
Radial profile of the momentum ratio (left) and energy ratio (right) of isothermal shells for strong radiation trapping vs. exponential and power-law luminosity decays: $L = 3 \times 10^{46}$erg/s, $f = 0.1$, $\tau_{IR,0} = 30$, $f_{dg} \sim 8 f_{\mathrm{dg,MW}}$ (red solid), $L_0 = 3 \times 10^{46}$erg/s, $t_c = 10^6$yr, $f = 0.005$, $\tau_{IR,0} = 1$, $f_{dg} \sim 6 f_{\mathrm{dg,MW}}$ (black dash-dot), $L_0 = 3 \times 10^{46}$erg/s, $\delta = 1$, $t_d = 10^6$yr, $f = 0.1$, $\tau_{IR,0} = 30$, $f_{dg} \sim 8 f_{\mathrm{dg,MW}}$ (green dotted). 
}
\label{plot_iso_var_traptime}
\end{figure*}

We have seen that in the case of fixed-mass shells, the high observed values of the outflow energetics may be attributed to either radiation trapping or luminosity decay. In fact, we have previously suggested that the observational measurements of the momentum boosts may be biased by AGN variability, but could not distinguish between the two possibilities \citep{Ishibashi_Fabian_2015}. 

Here we directly compare the effects of radiation trapping and luminosity decay on the inferred energetics of fixed-mass shells. 
In Fig. \ref{plot_VarTrap_fix_re}, we show the radial profiles of the momentum ratio and energy ratio for strong radiation trapping (red solid curve) and exponential luminosity decay (black dash-dot curve). 
In the case of radiation trapping, high values of the momentum and energy ratios are obtained at small radii, where the shell is optically thick to the reprocessed radiation (with the peak values reached close to the IR transparency radius); then $\zeta(r)$ and $\epsilon_k(r)$ decrease outwards with increasing radius. In contrast, in the case of the exponential luminosity decay, high values of the momentum and energy ratios are attained at large radii, with $\zeta(r)$ and $\epsilon_k(r)$ increasing outwards with increasing radius.

The distinct radial dependence of the outflow energetics may help distinguish between radiation trapping and luminosity decay.  
At present, most observational measurements report single global values of the outflow energetics at a given outflow radius \citep{Cicone_et_2014, Fiore_et_2017}.  
It should also be recalled that the outflow energetics parameters, defined in Eqs. (\ref{Eq_Mdot}-\ref{Eq_Ekdot}), are simple parametrisations of a time-dependent process.
In a few cases, the radial profiles of the outflow energetics have been obtained for a limited number of nearby sources. 
For instance, spatially resolved observations of Seyfert galaxies indicate that the mass outflow rate, momentum flux, and kinetic power, reach peak values at small radii and then steadily decrease outwards \citep{Crenshaw_et_2015, Revalski_et_2018}. In Mrk 231, the mass outflow rate, and to a lesser extent the kinetic power, are observed to decrease with increasing distance from the nucleus \citep{Feruglio_et_2015}. Such radially decreasing trends would be qualitatively consistent with the trapping of reprocessed radiation. 
We should however add a note of caution for Seyfert galaxies, which are usually moderate luminosity systems without extreme trapping conditions (although prominent dust lanes are observed in e.g. Mrk 573).
The radiation trapping scenario is particularly favoured for Mrk 231, as the source is currently accreting at a very high rate, close to the Eddington limit, hence the luminosity decay argument is unlikely to apply. 
Future observations providing detailed radial profiles of the outflow energetics should enable us to discriminate between different mechanisms. 

For completeness, Fig. \ref{plot_iso_var_traptime} shows an equivalent to Fig. \ref{plot_VarTrap_fix_re} for the case of expanding shells sweeping up mass (isothermal shells), coupled with different AGN luminosity decays. We observe that for strong radiation trapping, the maximal values of the momentum ratio and energy ratio are again attained at small radii, falling off with increasing radial distance (red solid curve).  High values of the outflow energetics may be obtained at large radii, in the case of an exponential luminosity decay coupled with a very low gas fraction (black dash-dot curve), as previously noted. 
On the other hand, a milder power-law luminosity decay cannot lead to high energetics values at large radii (green dotted curve), and only reach the peak on small scales due to radiation trapping. 
In the latter case, the radial profile of the outflow energetics becomes almost indistinguishable from the case of simple radiation trapping (with constant luminosity). 
The distinct radial dependence is thus most apparent in the case of fixed-mass shells and/or isothermal shells with very low gas fraction, coupled with a steep luminosity decay. 

We further note that for very low gas fractions ($f < 0.005$), even an increase in the dust-to-gas ratio, and hence IR opacity, cannot lead to significant radiation trapping. This result applies in general (as $\tau_{\mathrm{IR,0}} = \frac{\kappa_{IR} f \sigma^2}{2 \pi G R_0} \sim 6.4 \, \kappa_{\mathrm{IR}} f$, for fiducial parameters), regardless of the AGN luminosity decay history.


\subsection{The case of fossil outflows}

\begin{figure*}
\begin{multicols}{2}
    \includegraphics[width=0.8\linewidth]{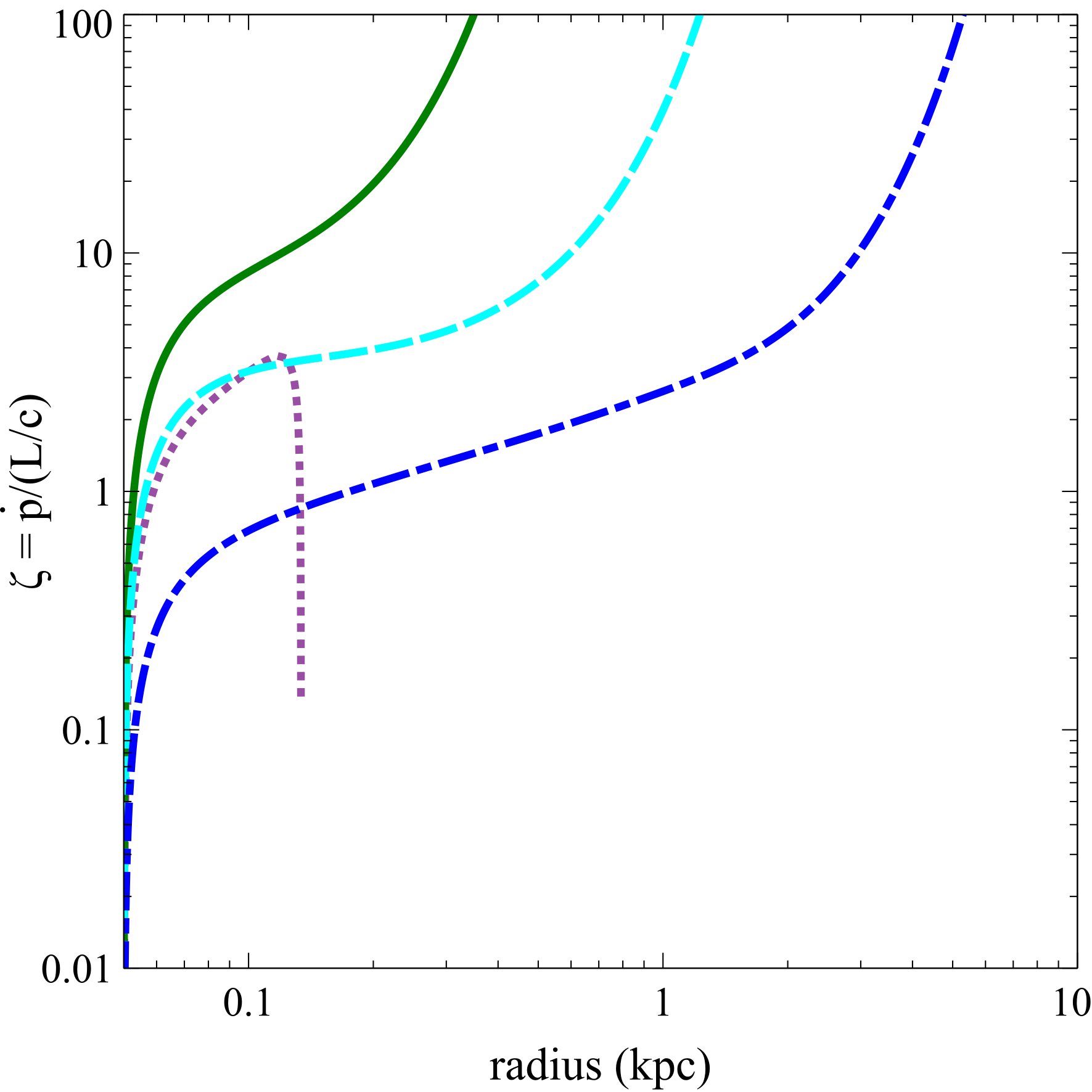}\par
    \includegraphics[width=0.8\linewidth]{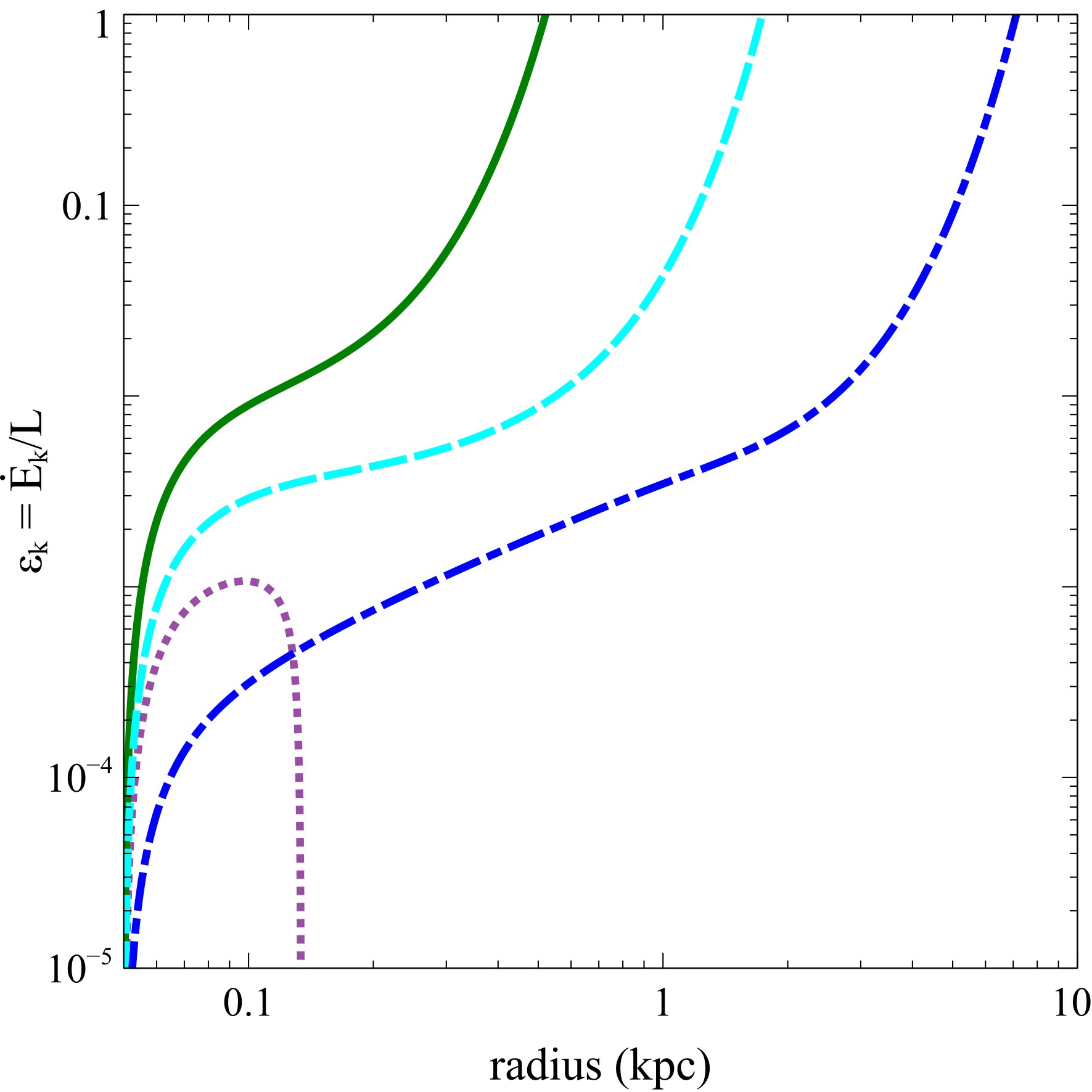}\par 
    \end{multicols}
\caption{
Radial profile of the momentum ratio (left) and energy ratio (right) of fixed-mass shells ($M_{sh}(r) = 10^8 M_{\odot})$ with exponential luminosity decay:  $t_c = 10^6$yr, $\tau_{IR,0}$ = 1, $f_{dg} \sim 1/500$ (blue dash-dotted); $t_c = 3 \times 10^5$yr, $\tau_{IR,0}$ = 5, $f_{dg} \sim 1/100$ (cyan dashed); $t_c = 10^5$yr, $\tau_{IR,0}$ = 10, $f_{dg} \sim 1/50$ (green solid). For comparison, a `failed' outflow with $t_c = 10^5$yr, $\tau_{IR,0}$ = 3.5, $f_{dg} \sim 1/150$ (violet dotted) is also shown. 
}
\label{plot_VarFossil_re}
\end{figure*}

Recent observations indicate the existence of outflows characterised by exceptionally high values of the momentum ratio ($\zeta > 100$) and energy ratio ($\epsilon_k > 0.1$) \citep{Fluetsch_et_2018}. These sources are clear outliers in the $\dot{p} - L/c$ and $\dot{E}_k - L$ relations, expected from the current luminosities, and represent good candidates for fossil outflows. We note that the extreme energetics of fossil outflows are often observed on small scales ($r \lesssim 1$kpc). 

In our picture, exceptionally high values of the momentum and energy ratios may be obtained in the case of fixed-mass shells coupled with an exponential luminosity decay; whereas such extreme values can never be achieved by radiation trapping alone (Fig. \ref{plot_VarTrap_fix_re}). 
In order to reproduce the extremely high values observed at small radii, very short decay timescales ($t_c < 10^6$yr) are required. In fact, a decrease in $t_c$ allows one to obtain higher energetics at smaller radii. But if the powering phase becomes too short, the outflow cannot propagate efficiently, and the shell will ultimately fall back. 
In such cases, a mild increase in radiation trapping allows one to have still shorter decay timescales. 
This combines to yield the highest energetics at small radii. Indeed, from Fig. \ref{plot_VarFossil_re} we see that very high values of $\zeta > 50$ and $\epsilon_k > 0.1$ can be obtained on small scales ($r \lesssim 1$kpc).
In contrast, such extreme values cannot be obtained by any form of radiation trapping. 
In Appendix \ref{Distinction_tauIR_tc}, we analyse more in detail the potential degeneracy between radiation trapping and luminosity decay. 
On the other hand, we note that a combination of very short decay timescale and modest radiation trapping leads to some form of `failed' outflow, which eventually falls back (violet dotted curve in Fig. \ref{plot_VarFossil_re}).

Fossil outflows with extreme energetics are observed in objects with very low Eddington ratios ($\lambda = L_{AGN}/L_E \sim 10^{-3}$, \citet{Fluetsch_et_2018}).  
In this context, we note that the inner accretion flow may possibly develop into some form of ADAF or RIAF, when the accretion rate drops below a critical value ($\dot{m} = \dot{M}/\dot{M}_E \sim 10^{-2}$) \citep[][and references therein]{Yuan_Narayan_2014}. 
In this regime, the radiative efficiency decreases rapidly (the radiative output having a quadratic dependence on the accretion rate), leading to a steep decline in AGN luminosity. Thus the driving UV luminosity may abruptly fall off, if the inner accretion disc transitions to an ADAF-like flow at low Eddington ratios.

Some of the fossil outflows, observed in galaxies with low AGN luminosity, are classified as purely star-forming galaxies \citep{Fluetsch_et_2018}. 
Observations of extreme molecular outflows in local ULIRGs, with low AGN fraction ($f_{AGN} \sim 0.2$), have been interpreted as requiring a combination of both AGN and starburst components \citep{Gowardhan_et_2018}. 
However, the lack of ongoing strong AGN activity should not rule out the possibility that the galactic outflows we observe at present were in fact driven by a past powerful AGN event that has since faded. For instance, the extreme outflow energetics could be explained by a rapid luminosity decay, coupled with modest radiation trapping (Fig. \ref{plot_VarFossil_re}). 
In this perspective, some of the powerful outflows observed on galactic scales, currently attributed to nuclear starbursts, could be re-interpreted in terms of previous AGN episodes. If this is the case, AGN feedback may play an even more prominent role in driving galactic outflows, with fossil outflows representing relics of past AGN history.

\begin{figure*}
\begin{multicols}{2}
    \hspace{0.7cm}
    \includegraphics[width=0.95\linewidth]{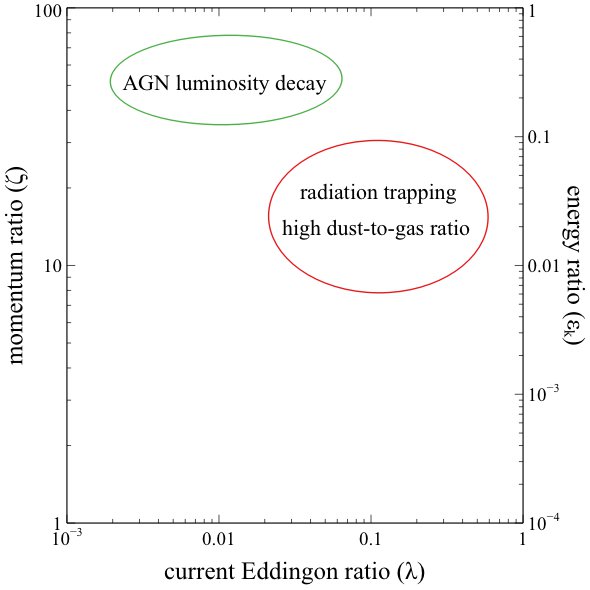}\par
    \hspace{0.7cm}
    \includegraphics[width=0.95\linewidth]{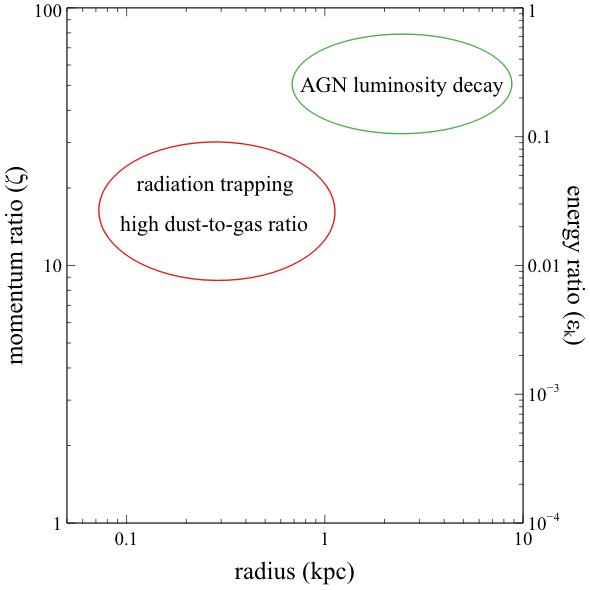}\par 
    \end{multicols}
\caption{A schematic diagram depicting the relative location of the dominant mechanism (AGN luminosity decay or radiation trapping) determining the observed outflow energetics. Note that the highlighted regions represent maximal values of the momentum and energy ratios. }
\label{plot_sketch}
\end{figure*}


\subsection{Physical implications}

The observed outflow properties may provide constraints on the past history of AGN activity and/or the physical conditions of the inner nucleus (see also the schematic diagram in Fig. \ref{plot_sketch}). 
In the case of fixed-mass shells, with negligible radiation trapping, the typically observed range of momentum and energy ratios seems to favour power-law luminosity decays; while the extreme energetics of fossil outflows may be explained by exponential decays. Moreover, observations of high energetics values at small radii ($r \lesssim 1$kpc) provides constraints on the characteristic decay timescales, which must be quite short ($< 10^6$yr). 
Based on simulations of the wind-driven outflow model, \citet{Zubovas_2018} argues that a power-law luminosity decay (with exponent $\sim 1$) adequately reproduces the observed range of momentum and energy loading factors, also preserving the correlations between outflow properties and AGN luminosity. 
Most recently, \citet{Nardini_Zubovas_2018} have further discussed how the multi-phase nature of wind-driven outflows may be used to study the accretion history of the central black hole. 
Therefore, if the observed outflow energetics can be attributed to AGN luminosity evolution, then one may potentially constrain the past history of AGN activity from the observed outflow properties. 

On the other hand, AGN luminosity evolution cannot account for the high energetics of expanding shells sweeping up ambient material, and radiation trapping is necessarily required. In these cases, the observed outflow properties may be used to put constraints on the physical conditions of the outflow launching region, such as the initial radius or the external density. 
We have previously tried a similar exercise for Mrk 231, and derived a rather small initial radius of $R_0 \sim 10$pc in this particular case \citep{Ishibashi_Fabian_2015}. 
Future observations providing the radial profiles of the outflow energetics will allow us to better constrain the physical nature of galactic outflows.


\section*{Acknowledgements }

ACF and WI acknowledge ERC Advanced Grant 340442.

  
\bibliographystyle{mn2e}
\bibliography{biblio.bib}


\appendix

\section{Dependence on $\MakeLowercase{n_0}$ and $R_0$}
\label{App_A1}

In Sect. \ref{Shell_mass_evolution}, we consider the energetics of expanding shells propagating in different ambient density distributions, characterised by different power-law exponents ($\alpha = 0, 1, 2$). Here we analyse the dependence on the external density ($n_0$) and the initial radius ($R_0$) in the case of $\alpha = 2$ shells. (We note that similar results are obtained when considering other density distributions.)

\begin{figure*}
\begin{multicols}{2}
    \includegraphics[width=0.7\linewidth]{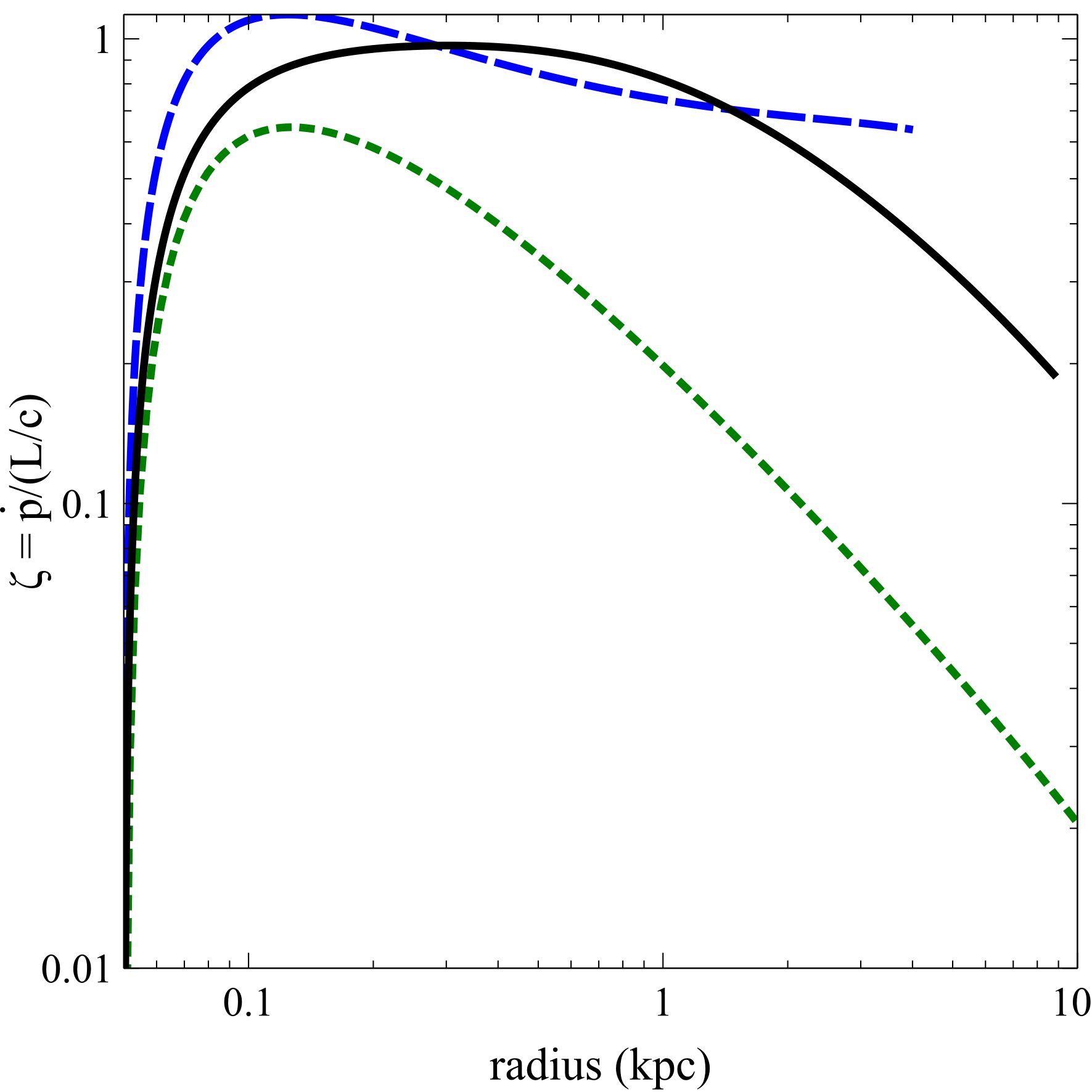}\par
    \includegraphics[width=0.7\linewidth]{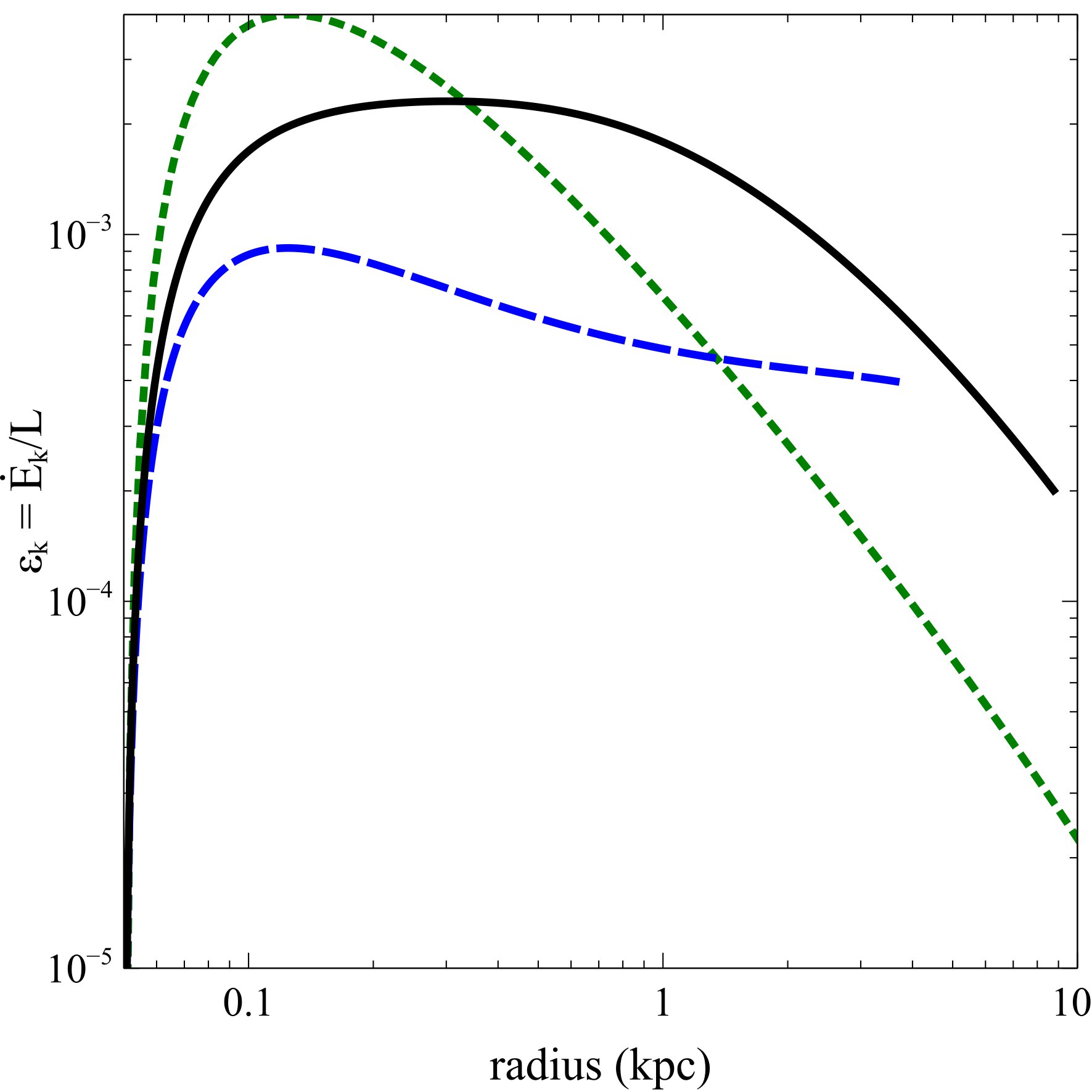}\par 
    \end{multicols}
\caption{
Radial profiles of the momentum ratio (left) and energy ratio (right) of $\alpha = 2$ shells with variations in the external density: $n_0 = 1000 cm^{-3}$ (blue dashed), $n_0 = 100 cm^{-3}$ (black solid), $n_0 = 10 cm^{-3}$ (green dotted).
}
\label{Fig_appendix_A1}
\end{figure*}

\begin{figure*}
\begin{multicols}{2}
    \includegraphics[width=0.7\linewidth]{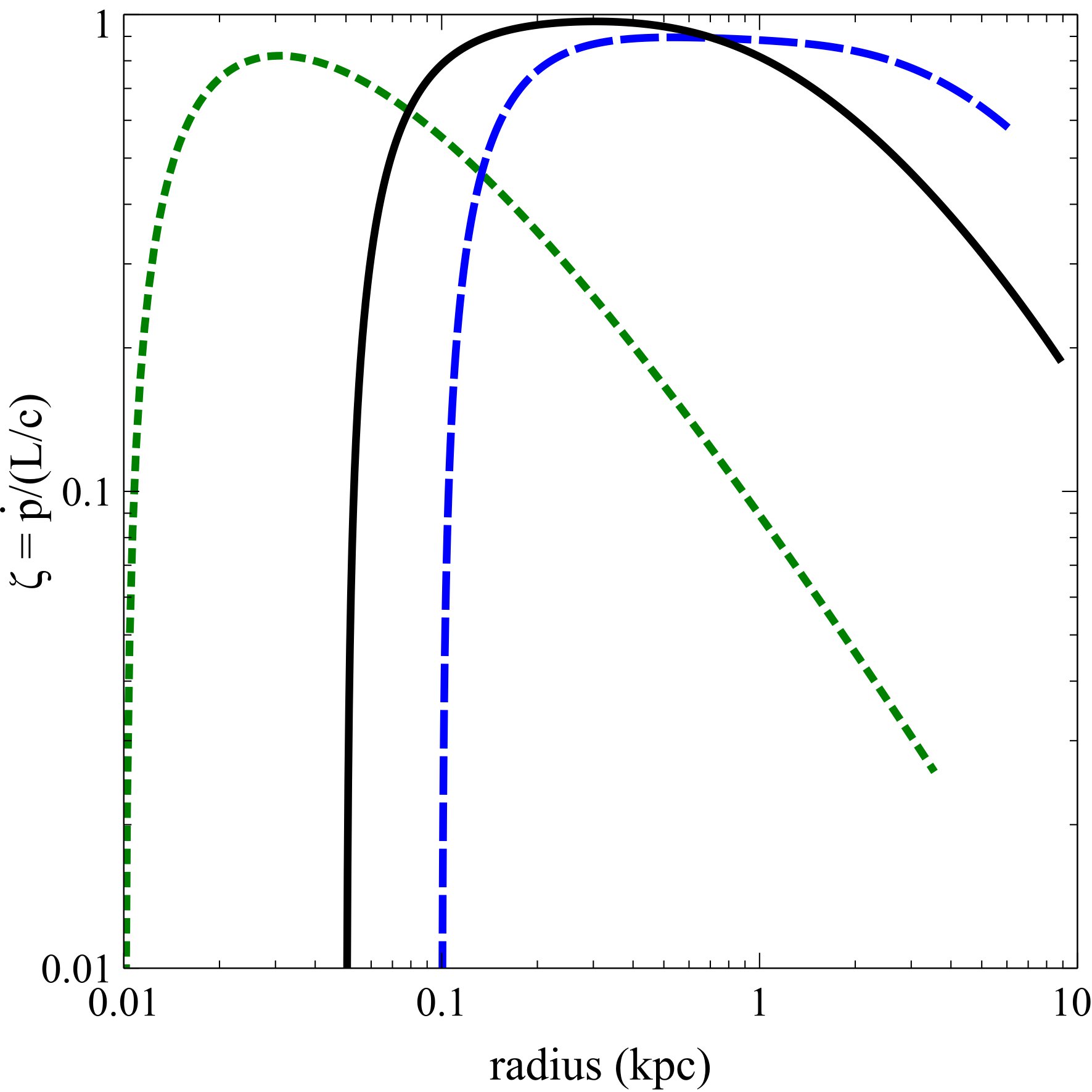}\par
    \includegraphics[width=0.7\linewidth]{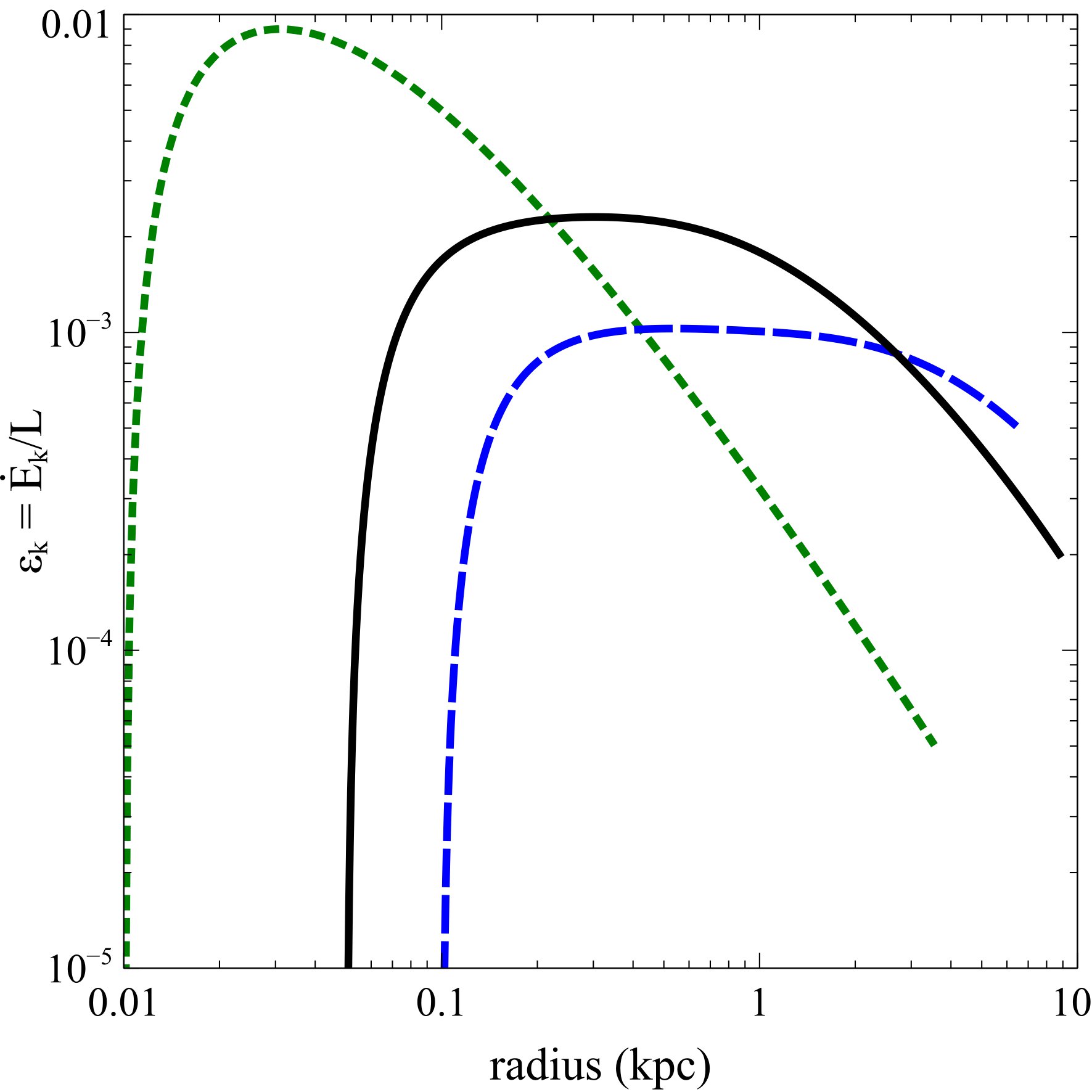}\par 
    \end{multicols}
\caption{
Radial profiles of the momentum ratio (left) and energy ratio (right) of $\alpha = 2$ shells with variations in the initial radius: $R_0 = 100$pc (blue dashed), $R_0 = 50$pc (black solid), $R_0 = 10$pc (green dotted). 
}
\label{Fig_appendix_A2}
\end{figure*}

From Fig. \ref{Fig_appendix_A1}, we see that the outflow energetics are not significantly affected by variations in the external density; in particular the momentum ratio always stays below $\zeta \lesssim 1$, and the energy ratio below a few times $\epsilon_k \sim 10^{-3}$. 
This trend was previously noticed in \citet{Ishibashi_Fabian_2015} for the case of the momentum ratio. 
Similarly, we observe that variations in the initial radius do not have a dramatic effect on the outflow energetics (Fig. \ref{Fig_appendix_A2}). In fact, a smaller initial radius implies a higher velocity, but also a lower mass outflow rate, such that the momentum ratios remain comparable and always below $\zeta \lesssim 1$. 
On the other hand, the velocity term is predominant in the energy ratio, and $\epsilon_k$ is slightly higher for smaller $R_0$, but never exceeds $ \sim 0.01$. 
These trends are in marked contrast to the case of fixed-mass shells, in which a decrease in the initial radius lead to a strong increase in the momentum ratio (cf. Fig. 6 in \citet{Ishibashi_Fabian_2015}).

We note that the slightly `convex' shape of the blue dashed curve in Fig. \ref{Fig_appendix_A1} is due to the fact that for larger $n_0$, the UV optical depth $\tau_{UV}$ is larger, and thus the shell becomes UV-optically thin at larger radii (the corresponding radial profiles of the radiative forces actually have distinct shapes).



\section{Distinction between large $\tau_{IR}$ and short $\MakeLowercase{t_c}$}
\label{Distinction_tauIR_tc}

\begin{figure*}
\begin{multicols}{2}
    \includegraphics[width=0.7\linewidth]{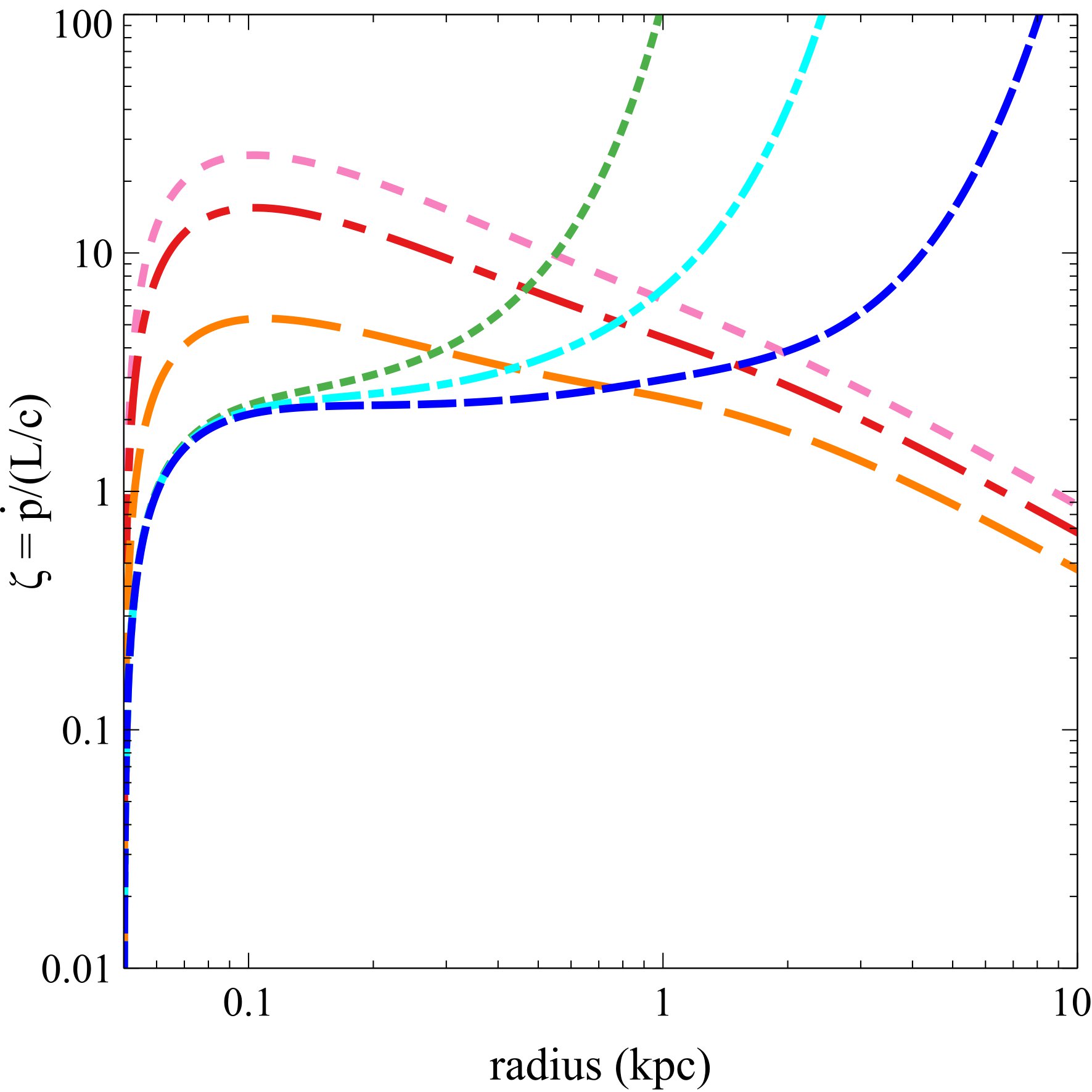}\par
    \includegraphics[width=0.7\linewidth]{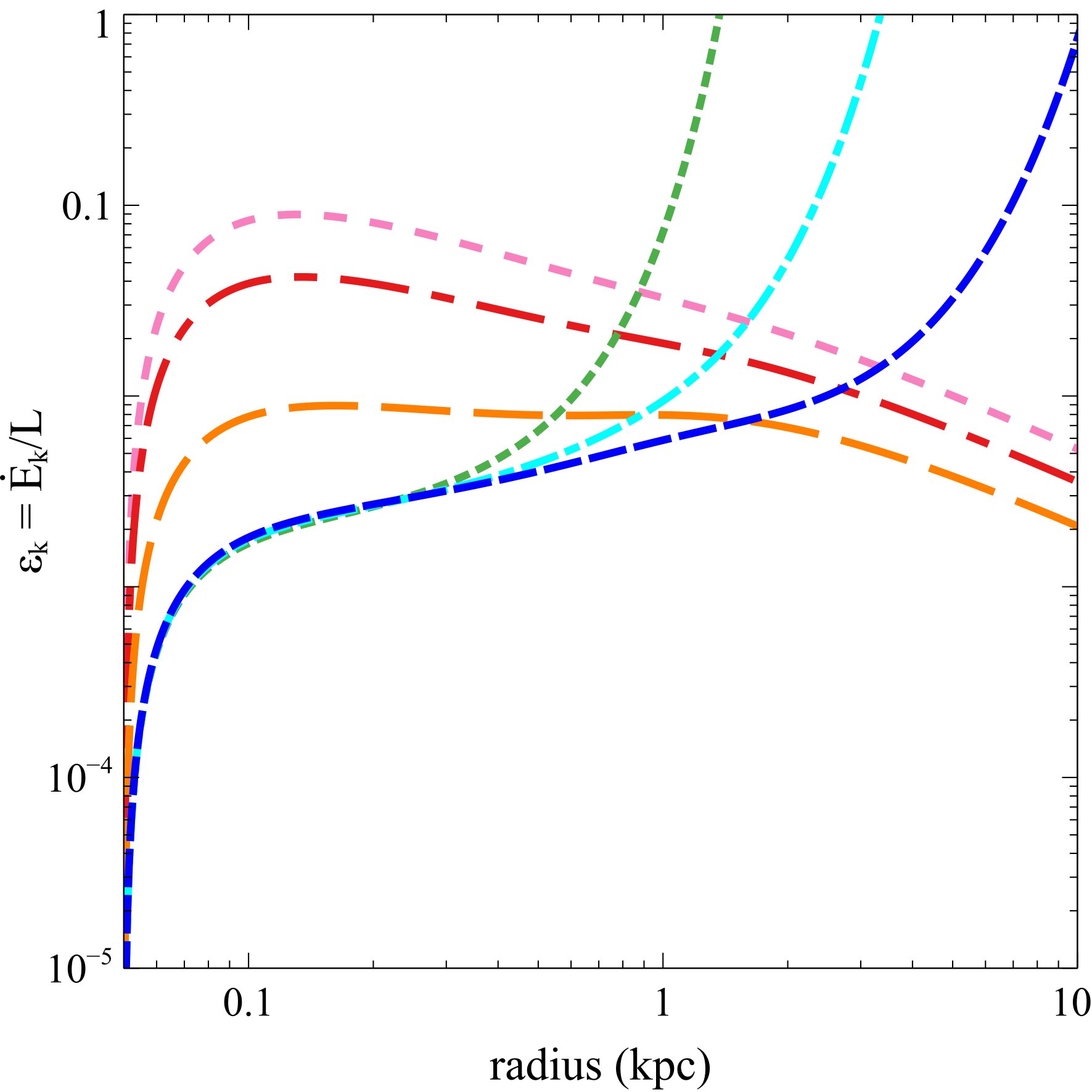}\par 
    \end{multicols}
\caption{Energetics of fixed-mass shells ($M_{sh} = 10^8 M_{\odot}$) with constant luminosity ($L = L_0 = 3 \times 10^{46}$erg/s) and different initial IR optical depth: $\tau_{IR,0} = 10$ (orange dashed-fine), $\tau_{IR,0} = 30$ (red dash-dot-fine), $\tau_{IR,0} = 50$ (pink dotted-fine); with exponential luminosity decay: $t_c = 10^6$yr (blue dashed), $t_c = 5 \times 10^5$yr (cyan dash-dot), $t_c = 3 \times 10^5$yr (green dotted). 
}
\label{plot_fix_shortTC_radtrap}
\end{figure*}

In Fig. \ref{plot_fix_shortTC_radtrap}, we plot the energetics of fixed-mass shells with different degrees of radiation trapping vs. different exponential luminosity decays. 
An increase in the IR optical depth leads to higher values of the momentum ratio and energy ratio at small radii (with the peak values reached around the same radius), and $\zeta(r)$ and $\epsilon_k(r)$ decreasing with increasing radius. On the other hand, a decrease in the characteristic decay timescale leads to higher values of the energetics at smaller radii, with $\zeta(r)$ and $\epsilon_k(r)$ increasing with radial distance.
But if the powering phase becomes too short, the outflow cannot propagate anymore, and this sets a lower limit on $t_c$. 
As discussed in Sect. \ref{discussion}, radiation trapping may account for moderately high values of the outflow energetics, while the extreme values observed in e.g. fossil outflows may only be accounted for by exponential luminosity decays. 
We note that there is some overlap at intermediate radii ($r \sim 1$kpc), where it will be difficult to discriminate between radiation trapping and luminosity decay. In principle, the distinct radial dependence of $\zeta(r)$ and $\epsilon_k(r)$ should help distinguish between the two possibilities, if multiple measurements of the outflow energetics are available at different radii.

\label{lastpage}

\end{document}